\documentclass[]{article}%

\usepackage{graphicx}
\usepackage{multicol,multirow}
\usepackage{amsmath,amssymb,amsfonts}
\usepackage{mathrsfs}
\usepackage{amsthm}
\usepackage{rotating}
\usepackage{appendix}
\usepackage{ifpdf}
\usepackage[T1]{fontenc}
\usepackage{newtxtext}
\usepackage{newtxmath}
\usepackage{textcomp}
\usepackage{xcolor}
\usepackage{lipsum}
\usepackage[colorlinks,allcolors=blue]{hyperref}
\usepackage{anyfontsize}
\usepackage{enumerate}
\usepackage{enumitem}

\usepackage[a4paper, margin=2.5cm]{geometry}

\theoremstyle{definition}

\numberwithin{equation}{section}

\usepackage{natbib}
\usepackage{booktabs}

\title{Fuzzy entropic triple $k$‑means for simultaneous three way clustering}

\author{Mariaelena Bottazzi Schenone, Roberto Rocci,Maurizio Vichi}

\date{Department of Statistical Sciences, Sapienza University of Rome, Rome, Italy}

\begin{document}

\maketitle

\abstract{This paper proposes fuzzy entropic triple $k$-means (FE3KM), an entropy-regularized fuzzy partitioning method for three-way data arrays that simultaneously clusters objects, variables, and occasions. 
The method combines three row-stochastic fuzzy membership matrices and a full array of block centroids through a product-weighted squared-distance criterion with mode-specific entropy regularization to represent uncertainty in cluster assignments.
With normalized product membership weights and weighted-mean block centroids, the ordinary total sum of squares decomposes exactly into within- and between-block components, further attributable to each mode and to individual clusters within each. 
This identity supports cluster-level descriptive summaries and conditional assessments of the gain associated with each way's partition.
We derive a block-coordinate descent algorithm with weighted-mean centroid updates and entropy-based membership updates, and clarify the relationship with tandem procedures, independent partitioning, hard three-way partitioning, and Tucker reconstruction. An accelerated variant is also proposed for large arrays.
A simulation study examines partition and centroid recovery under different noise levels and array sizes, sensitivity to the entropy regularization parameters, and a comparison with tandem and independent procedures.
An application to a benchmark three-way TV ratings dataset shows how FE3KM illustrates interpretable object-variable-occasion structures and quantifies the contribution of each cluster and mode to overall data variability.}




\section{Introduction}\label{sec:introduction}

Three-way data arise when a set of units is described by multiple variables observed on different occasions or under different conditions. This structure is particularly useful for studying complex phenomena whose multivariate characteristics vary across time, contexts, or experimental settings. Such data take the form of a three-way array $\mathbf{\underline{X}} = [x_{ijk}]$ of $IJK$ values, referred to $J$ variables measured on $I$ objects at $K$ occasions \citep{coppi1989multiway, kroonenberg2008applied}. 
The term way denotes a dimension of the data, whereas mode refers to its substantive interpretation. Preserving this structure allows the investigation of the relationships among units, variables, and occasions jointly, rather than examining each occasion separately or combining dimensions in ways that obscure their distinct roles.
Three-way structures occur naturally in psychometric research. In longitudinal studies of psychological well-being, individuals may complete several symptom or functioning scales at successive assessment waves, producing a person-by-measure-by-occasion array. Educational studies may record students’ performance across cognitive tasks or skill domains at different stages of learning. Similarly, personality research may assess individuals on multiple traits across situations, while intervention studies may measure several psychological outcomes before, during, and after treatment. Related structures are common in economics and the social sciences: countries can be described by economic indicators over several years, households by expenditure categories across survey waves, and geographical areas by social indicators across successive observation periods. In each case, the analytical challenge is to summarize heterogeneity while retaining the connections among the three modes.
Clustering provides a useful approach to this challenge by identifying groups with similar profiles and reducing a large array to a more interpretable representation. For three-way data, the simultaneous clustering of units, variables, and occasions can reveal groups of units characterized by particular sets of attributes under particular temporal or contextual conditions. 

Several strategies have been proposed to this end. They can be grouped according to what is clustered and how. (\emph{a}) \emph{Least-squares partitioning of all three
ways}: the array is approximated by block centroids defined through three partitions of objects, variables and occasions \citep{rocci2003threemode,kiers2004,schepers06}.
(\emph{b}) \emph{Component analysis combined with a partition of the units}: a crisp or
fuzzy partition of one mode is estimated together with components for the other ways
\citep{Rocci2005,Vichi2007}. (\emph{c}) \emph{Factorization followed by
clustering}: a CANDECOMP/PARAFAC or Tucker factorization is estimated first, and $k$-means or related methods are applied in the factor space
\citep{huang08,sun16,wu16}. (\emph{d}) \emph{Tensor block models and spectral or tri-clustering algorithms}, which estimate block structures through generative or
spectral criteria \citep{wang2026incremental,battaglia23,feizi17,andriantsiory21}; see
\citet{henriques18} and \citet{swathypriyadharsini24} for comprehensive reviews.
(\emph{e}) \emph{Probabilistic mixture models} for three-way data \citep{Vermunt2007}.
These methods differ in their objective functions (Least-Squares, spectral or probabilistic), in the constraints imposed on the factor or membership matrices, and in
whether the three modes are clustered sequentially, as in tandem procedures, or simultaneously.

A separate line of research, dating back to Ruspini's formulation of fuzzy partitions
\citep{ruspini1969new} and Dunn's fuzzy relative of ISODATA \citep{dunn1973fuzzy}, has shown
that fuzzy memberships offer a more realistic representation of cluster structure than
hard partitions whenever boundaries are not sharp \citep{hoppner1999fuzzy}. Intermediate
memberships are particularly valuable in applications such as customer segmentation,
where they carry diagnostic value for transitional or anomalous entities that a hard
partition would otherwise obscure \citep{durso2012wavelet,ruspini2019fuzzy}, making fuzzy
memberships a genuine source of information rather than a mere modeling convenience.
Two main methodologies are used for fuzzy clustering of objects: fuzzy $c$-means \citep{Bezdek1981} and fuzzy entropy-regularized $k$-means \citep{liu15}. The first controls membership diffuseness through an exponent applied to membership weights, whereas the second combines a linearly membership-weighted squared-distance loss with an entropy regularization term that encourages diffuse assignments. Fuzzy memberships have been introduced in three-way clustering in several ways: a crisp or fuzzy partition of the units in component-analysis models \citep{Rocci2005}, fuzzy clustering for
longitudinal three-way data \citep{coppi10}, fuzzy triclustering of three-dimensional cubes \citep{liu15}, three-mode fuzzy co-clustering of co-occurrence data \citep{honda21}, and, in the statistical literature, tensor mixed-membership blockmodels \citep{Agterberg2025}. Despite this, fuzzy memberships have rarely been extended to all modes of a three-way array. 

Naive alternatives, i.e., applying separate one-mode fuzzy clustering to each dimension, ignore cross-mode interactions and typically yield sub-optimal block structures. Recent developments in multi-view fuzzy clustering and in incremental fuzzy models for evolving data \citep{sun2026multiview, wang2026incremental} further motivate fuzzy formulations able to jointly capture uncertainty and heterogeneity across modes.

Entropy regularization is also used in the three-way fuzzy literature: the fuzzy tri-clustering algorithm of \cite{liu15} defines a membership function for each of
the three dimensions and combines an aggregation term with one entropy term per dimension. Entropy regularization is moreover a well-established fuzziness-control mechanism in other complex-data settings, from its original least-squares formulation for multivariate time trajectories \citep{coppi2006fuzzy} to robust entropy-based clustering of time series \citep{durso2023robust} and interval-valued time series
\citep{vitale2026entropy}.

In this spirit, we propose here fuzzy entropic triple $k$-means (FE3KM), a new methodology for jointly partitioning the three modes of a numerical data array. The method uses three separately indexed (fuzzy) row-stochastic membership matrices and a full array of block centroids. FE3KM loss function combines a product-weighted sum of squared distances to these centroids with mode-specific entropy regularization, allowing the degree of membership fuzziness to differ across modes.  What distinguishes FE3KM is therefore not the use of entropy but the form of the criterion and of the partition structure: FE3KM minimizes a product-weighted least-squares distance to a full $P\times Q\times R$ array of block centroids, with three separately indexed row-stochastic partitions.

In FE3KM, the entropic approach preserves the normalization of product memberships and, with the corresponding weighted-mean centroids, permits an exact decomposition of the ordinary total sum of squares into within-block and between-block components. This decomposition provides a common scale for summarizing between-block variation and for assessing the conditional gain associated with retaining a partition in one mode while holding the other two memberships fixed. Such summaries describe the optimal fitted solution. Moreover, positive entropy regularization yields well-defined softmax membership updates, and allows the degree of diffuseness to be controlled separately across modes. These are interpretive and computational advantages. The computational advantages are related to the fact that the positive entropy regularization yields explicit normalized exponential membership updates that remain well-defined in the presence of zero distances and preserve positive cluster masses (non-empty clusters) in exact arithmetic. These properties also simplify the implementation. 

FE3KM can be viewed as a fuzzy extension of triple $k$-means \citep{rocci2003threemode} and three-mode partitioning \citep{schepers06}. It has the form of a Tucker3 model, $\hat x_{ijk}=\sum_{p,q,r}u_{ip}v_{jq}w_{kr}g_{pqr}$,
in which the component matrices are row-stochastic membership matrices and the core contains block centroids. However, its criterion is not the Tucker reconstruction error: Proposition~1 will show that it equals that error plus a nonnegative centroid-dispersion term. Unlike standard fuzzy clustering based on the exponent $m$ (e.g., fuzzy $c$-means),where memberships enter the objective nonlinearly, FE3KM employs linear memberships within a Least-Squares objective function regularized by entropy. The joint estimation is motivated by the interdependence of the three partitions. Unlike separate or sequential analyses, it repeatedly revises each partition using the current structure in the other modes. This permits a unified description of cross-mode response patterns.

FE3KM offers a cluster-level interpretation based on an exact decomposition of the Total Sum of Squares, which holds because the product membership weights are normalized and the block centroids are weighted means. Similarly, several clustering algorithms are formulated purely as within-cluster minimization problems, treating
between-cluster separation as an implicit, post-hoc diagnostic rather than a quantity the optimization explicitly controls. As shown in Subsection \ref{devdec}, FE3KM closes this gap: the product weights $u_{ip}v_{jq}w_{kr}$ of row-stochastic memberships sum to one over the blocks, so that the criterion is a weighted version of the ordinary Total Sum of Squares and minimizing the within-cluster deviance is exactly equivalent to maximizing the between-cluster
sum of squares (deviance), up to the entropy penalty terms (Remark~3), in the classical sense of \cite{cormack1971review}, so that separation
between blocks is a built-in property of the objective rather than an afterthought.

Building on these premises, the paper makes
three main contributions. First, it develops an entropy-regularized, product-weighted Least-Squares formulation with separately indexed fuzzy partitions for the three modes and weighted-mean block centroids. Its relationship to hard three-mode partitioning, fuzzy triclustering, and Tucker reconstruction is made explicit. Second, it derives a block-coordinate descent algorithm (standard and accelerated) consistent with this criterion and clarifies its relationship to sequential clustering procedures. Third, it establishes the conditions for an ordinary total sum-of-squares decomposition, ANOVA-like, and develops cluster-level descriptive summaries and conditional mode-pooling contrasts. These results are intended to support interpretation of the fitted block structure,
providing an interpretable framework for three-way fuzzy clustering. 

The remainder of the paper is organized as follows. Section \ref{sec2} lists the notation used throughout. Section \ref{sec3} discusses three-way partitioning in
detail: Subsection
\ref{sec:simultaneous} presents the simultaneous FE3KM approach; while Subsection \ref{sec:nonsim} introduces the non-simultaneous tandem and independent procedures, comparing them with the proposed model; Subsection \ref{sec3.4} details the Alternating Least-Squares (ALS)
algorithm, including an accelerated version for larger arrays; Subsection \ref{devdec} derives the deviance decompositions; and Subsection \ref{compare}
positions FE3KM relative to alternative three-way clustering methods. The proposed methodology is then evaluated on synthetic and real data in Sections \ref{sec4} and \ref{sec5}, respectively, with concluding remarks in Section \ref{sec6}.

\section{Notation}\label{sec2}

For the convenience of the reader, the notation and terminology common to all sections is listed here:

\begin{itemize}[label = -, itemsep=0cm]
    \item $I, J, K, P, Q, R$: number of units, variables, occasions, clusters of objects, clusters of variables, and clusters of occasions, respectively;    
    \item $\textbf{\underline{X}},\ \textbf{\underline{E}}\ \ (I \times J \times K)$ three-way data array and three-way array of error terms, where $x_{ijk}$ ($e_{ijk}$) is the value (error) of the $j$th variable on the $i$th object at the $k$th occasion;    
    \item $\mathbf{X}_{I,JK} = \left[\mathbf{X}_{..1},\ldots,\mathbf{X}_{..K}\right]$,  $(I \times JK)$ matricized data array $\underline{\mathbf{X}}$ obtained by placing its frontal slabs side by side; 
    \item $\mathbf{X}_{J,IK} = \left[\mathbf{X}_{1..},\ldots,\mathbf{X}_{I..}\right]$, $(J \times IK)$ matricized data array $\underline{\mathbf{X}}$ obtained by placing its flat slabs side by side;
    \item $\mathbf{X}_{K,IJ} = \left[\mathbf{X}_{.1.},\ldots,\mathbf{X}_{.J.}\right]$, $(K \times IJ)$ matricized data array $\underline{\mathbf{X}}$ obtained by placing its lateral slabs side by side;
    \item $\mathbf{E}_{I,JK} = [\mathbf{E}_{..1}, \dots, \mathbf{E}_{..K}]$, $(I \times JK)$ matricized errors array $\underline{\mathbf{E}}$ obtained by placing its frontal slabs side by side;   
    \item $\mathbf{U} = [u_{ip}]$ $(I \times P)$ membership matrix defining a fuzzy partition of objects into $P$ clusters, where $u_{ip} \in [0,1]$ is the degree of membership of object $i$ to cluster $p$. Matrix $\mathbf{U}$ is row stochastic;
    \item $\mathbf{V} = [v_{jq}]$ $(J \times Q)$ membership matrix defining a fuzzy partition of variables into $Q$ clusters, where $v_{jq} \in [0,1]$ is the degree of membership of variable $j$ to cluster $q$. Matrix $\mathbf{V}$ is row stochastic;
    \item $\mathbf{W} = [w_{kr}]$ $(K \times R)$ membership matrix defining a fuzzy partition of occasions into $R$ clusters, where $w_{kr} \in [0,1]$ is the degree of membership of occasion $k$ to cluster $r$. Matrix $\mathbf{W}$ is row stochastic;
    \item $\lambda_\mathbf{U}, \lambda_\mathbf{V}, \lambda_\mathbf{W}$: penalties for entropic fuzzy partitioning;
    \item $\|\underline{\mathbf A}\|$: Euclidean (Frobenius) norm of an array, defined by
$\|\underline{\mathbf A}\|^2=\sum a_{ijk}^2$, the sum of the squares of its entries (for a matrix or a vector, the same sum over its own indices);
    \item $\mathbf{I}_a$ $(a \times a)$ identity matrix;
    \item diag($\mathbf{a}$) ($a \times a)$ diagonal matrix with vector $\mathbf{a}$ on the diagonal.
\end{itemize}

\section{Fuzzy entropic partitioning of a three-way dataset}\label{sec3}

\subsection{Simultaneous model formulation}\label{sec:simultaneous}
Let us assume to have data on $J$ variables measured on $I$ objects at $K$ different occasions, organized in a three-way array $\mathbf{\underline{X}} = [x_{ijk}]$. 
We seek P object clusters, Q variable clusters, and R occasion clusters.

The three partitions are represented by membership matrices $U = [\mathbf{U}_{ip}], \mathbf{V} = [v_{jq}]$, and $\mathbf{W} = [w_{kr}]$, satisfying: 
$\mathbf{U1}_P=\mathbf{1}_I$, 
$\mathbf{V1}_Q=\mathbf{1}_J$, 
$\mathbf{W1}_R=\mathbf{1}_K$, 
i.e., row-stochastic.
Thus, each entity has a membership vector on the corresponding probability simplex. These vectors describe fuzzy cluster association. They are without a probabilistic interpretation and do not measure the sampling uncertainty, but describe the classification uncertainty of objects, variables, and occasions.

Let $\underline{\mathbf{G}} = [g_{pqr}]$ be a ($P \times Q \times R$) array of block centroids. The product: 
$a_{ijk,pqr} = u_{ip}v_{jq}w_{kr}$
measures the association of observation $x_{ijk}$ with block $(p,q,r)$. The row-stochastic constraints imply that $\sum_{pqr} a_{ijk,pqr} = 1$ for every observation $x_{ijk}$. In fact, $\sum_p \sum_q \sum_r a_{ijk,pqr}
= \sum_p \sum_q \sum_r u_{ip} v_{jq} w_{kr}
= (\sum_p u_{ip})(\sum_q v_{jq})(\sum_r w_{kr})
= 1 \times 1 \times 1
= 1.$
We are now in position to state the Fuzzy Entropic triple $k$-means model, briefly FE3KM. It minimizes the following objective function
\begin{align}
    & f_{FE3KM}(\mathbf{U},\mathbf{V},\mathbf{W},\underline{\mathbf{G}}) = L(\mathbf{U},\mathbf{V},\mathbf{W},\underline{\mathbf{G}}) + \lambda_\mathbf{U}   {p}(\mathbf{U}) + \lambda_\mathbf{V}   {p}(\mathbf{V}) + \lambda_\mathbf{W}   {p}(\mathbf{W}). 
\end{align}\label{obj_fe3km}
where
\begin{align}\label{eq_3.2}
L(\mathbf{U},\mathbf{V},\mathbf{W},\underline{\mathbf{G}}) = \sum_{i=1}^{I} \sum_{j=1}^{J} \sum_{k=1}^{K} \sum_{p=1}^{P} \sum_{q=1}^{Q} \sum_{r=1}^{R} u_{ip} v_{jq} w_{kr} \left( x_{ijk} - g_{pqr} \right)^2 
\end{align}
is the unregularized weighted Least-Squares (LS) loss of triple $k$-means and 
\begin{align}\label{eq_8}
{p}(\mathbf{U}) = \sum_{i=1}^I\sum_{p=1}^Pu_{ip}\log(u_{ip}), \\
{p}(\mathbf{V}) = \sum_{j=1}^J\sum_{q=1}^Qv_{jq}\log(v_{jq}), \\
{p}(\mathbf{W}) = \sum_{k=1}^K\sum_{r=1}^Rw_{kr}\log(w_{kr}).
\end{align}
are the penalty terms, while  $\lambda_\mathbf U, \lambda_\mathbf V$, and $\lambda_\mathbf W$ are strictly positive constants.

By using this loss function, we require that the blocks must consist of entries that are as similar as possible in a LS sense. 

The squared Euclidean distance in (\ref{obj_fe3km}) is chosen for four reasons: ($i$) it extends the classical $k$-means objective to the three-way fuzzy setting, ($ii$) it yields closed-form centroid updates as weighted averages, ($iii$) it ensures convexity in $\underline{\mathbf{G}}$ when memberships are fixed, and ($iv$) it is the standard discrepancy in fuzzy clustering, ensuring comparability with existing methods.

While squared Euclidean is adopted here, alternative measures may be preferable for certain data types. Absolute deviation ($L_1$) offers robustness to outliers but requires solving weighted median problems. Divergence measures (e.g., Kullback-Leibler, Itakura-Saito) are suitable for count data. More generally, Minkowski distances of order $p$ allow tuning sensitivity to outliers. However, squared Euclidean remains the most tractable and widely used in fuzzy clustering.

The entropy penalties $\lambda_\mathbf{U} p(\mathbf{U})$, $\lambda_\mathbf{V} p(\mathbf{V})$, $\lambda_\mathbf{W} p(\mathbf{W})$ act as fuzziness controllers. As positive regularization coefficients approach zero, conditional membership updates approach hard assignments when the minimizing distances are unique, and memberships tend to be hard (crisp); as $\lambda$ increases, memberships approach uniformity (maximally fuzzy). This allows the model to represent uncertainty in noisy data, prevent overconfident assignments, and avoid degenerate solutions. The $\lambda$'s are selected data-adaptively via cross-validation (Section \ref{sec3.4}).

This model is called \lq\lq \textit{fuzzy entropic triple $k$-means}\rq\rq\ (FE3KM) because it can be considered the three-way fuzzy entropic partitioning version of \cite{rocci2003threemode}, that is the three-way version of the standard $k$-means algorithm \citep{MacQueen1967}. The loss in (\ref{obj_fe3km}) has several interesting particular cases. For example, when the three lambda coefficients are equal to zero and the number of clusters of variables and occasions is equal to the number of variables and occasions, respectively, then the proposed approach reduces to the $k$-means algorithm applied on the matricized array $\mathbf{X}_{I,JK}$ for classifying objects according to variables and occasions. Moreover, when variables and objects are observed at only one occasion ($K = 1$) and the penalty terms are again equal to zero, then (\ref{obj_fe3km}) reduces to the double $k$-means of \cite{Vichi2000}.

When the penalty parameters are set to zero, the loss function in \eqref{obj_fe3km} reduces to a clustering-oriented version of the Tucker3 model \citep{Tucker1966}. In Tucker3, a three-way array $X$ is approximated as $x_{ijk} \approx \sum_{p=1}^P \sum_{q=1}^Q \sum_{r=1}^R a_{ip} b_{jq} c_{kr} g_{pqr}$, where $A,B,C$ are component matrices (often orthonormal) and $G$ is a core array. 
FE3KM defines a Tucker-form reconstructed array through three row-stochastic membership matrices and a full core. Its estimation criterion, however, is a product-weighted distance loss rather than ordinary Tucker reconstruction error. Proposition 1 shows that these criteria differ by a nonnegative centroid-dispersion term.
This connects FE3KM to a well-established family of multiway factorization models while emphasizing its role as a fuzzy three-way partitioning method.
This connection is important because it links our clustering approach to a well-established family of dimensionality reduction methods, highlighting how FE3KM simultaneously performs dimensionality reduction (via the core $\underline{\mathbf{G}}$) and fuzzy clustering (via $\mathbf{U}$, $\mathbf{V}$, $\mathbf{W}$). \cite{schepers06} discussed several algorithms for fitting loss (\ref{eq_3.2}) when the lambdas are zero. 

\vspace{0.3cm} \noindent
\textbf{Remark 1}: The membership matrices and core array define a reconstructed array $\hat{x}_{ijk}$ with entries: $\hat{x}_{ijk} = \sum_{p=1}^{P}\sum_{q=1}^{Q}\sum_{r=1}^{R} u_{ip}v_{jq}w_{kr}g_{pqr}.$
These are the weighted averages of the block centroids.

\vspace{0.3cm} \noindent
\textbf{Proposition 1}: Function $L(\mathbf{U},\mathbf{V},\mathbf{W},\underline{\mathbf{G}})$ represents reconstruction error plus the weighted dispersion of the block centroids. For row-stochastic $\mathbf{U}$, $\mathbf{V}$, and $\mathbf{W}$ and any core $\underline{\mathbf{G}}$,

$L(\mathbf{U},\mathbf{V},\mathbf{W},\underline{\mathbf{G}}) = \sum_{ijk} (x_{ijk} - \hat{x}_{ijk})^2
  + \sum_{ijk} \sum_{pqr} a_{ijk,pqr} (g_{pqr} - \hat{x}_{ijk})^2$

\vspace{0.2cm}
\noindent
\textbf{Proof}. For each observation, write $x_{ijk} - g_{pqr}$ as the sum of $x_{ijk} - \hat{x}_{ijk}$ and $\hat{x}_{ijk} - g_{pqr}$. After squaring and summing with weights $a_{ijk,pqr}$, the cross-term vanishes because the weights sum to one and their weighted mean of $g_{pqr}$ is $\hat{x}_{ijk}$. Summation over observations gives Equation ($\ref{eq_3.2}$).

\vspace{0.2cm}
The second term of $L(\mathbf{U},\mathbf{V},\mathbf{W},\underline{\mathbf{G}})$ is nonnegative and measures the dispersion of block centroids receiving positive weight for an observation. Hard memberships make this term zero. It is zero for hard memberships, and the dispersion term is zero if, for every observation, all block centroids receiving positive membership weight have the same value.

Proposition 1 shows that minimizing $L(\mathbf{U},\mathbf{V},\mathbf{W},\underline{\mathbf{G}})$ balances accurate reconstruction with low dispersion among the block centroids contributing to each fitted entry. Thus, $L(\mathbf{U},\mathbf{V},\mathbf{W},\underline{\mathbf{G}})$ penalizes membership assignments that combine widely differing centroids. The resulting reconstruction can be used to estimate missing entries through $\hat{x}_{ijk}$, provided that the model is fitted using only observed entries.

\subsection{Non-simultaneous procedures for three-way fuzzy partitioning}\label{sec:nonsim}

In this section, some alternative procedures for fuzzy partitioning the three-way data matrix $\mathbf{\underline{X}}$ are considered. They can be seen as a multi-way partitioning obtained by applying fuzzy entropic $k$-means on each way, either sequentially (and for this reasons they are briefly called \lq\lq tandem procedures\rq\rq, Section \ref{tandem}) or independently (Section \ref{sec:independent}).

\subsubsection{Tandem procedures}\label{tandem}


Tandem procedures \citep{arabie1990three} for clustering objects, variables, and occasions may be defined by means of classical clustering methodologies applied sequentially in six different ways: 1) obj $\Rightarrow$ var $\Rightarrow$ occ; 2) obj $\Rightarrow$ occ $\Rightarrow$ var; 3) var $\Rightarrow$ obj $\Rightarrow$ occ; 4) var $\Rightarrow$ occ $\Rightarrow$ obj; 5) occ $\Rightarrow$ obj $\Rightarrow$ var; 6) occ $\Rightarrow$ var $\Rightarrow$ obj. Only the first sequential procedure will be shown, that is, obj $\Rightarrow$ var $\Rightarrow$ occ, leaving out all the other permutations for the sake of space, since they can be similarly derived.

The procedure starts by partitioning the objects according to the fuzzy entropic $k$-means algorithm applied to $\underline{\mathbf{X}}$, so as to obtain the object membership matrix $\mathbf{U}$ and the object centroids ($P \times J \times K)$ array $\underline{\mathbf{M}}$. The procedure continues by partitioning variables using the object centroids weighted by the cardinalities of the object clusters. In this way, a variable membership matrix $\mathbf{V}$ is obtained, along with a object-variable centroids ($P \times Q \times K)$ array $\underline{\mathbf{C}}$. The procedure ends with the partition of the occasions, applying again fuzzy entropic $k$-means to the object-variable centroid array $\underline{\mathbf{C}}$, weighted by the cardinalities of the object and variable clusters, to obtain an occasion membership matrix $\mathbf{W}$ and an object-variable-occasion centroid array $\underline{\mathbf{G}}$ of dimension ($P \times Q \times R$). This procedure can be formally written as follows.

\begin{enumerate}
    \item[a.] \textbf{Partitioning of objects} via fuzzy entropic $k$-means algorithm applied to $\underline{\mathbf{X}}$ by minimizing
    \begin{equation}\label{1}
       f_1(\mathbf{U},\underline{\mathbf{M}}) = \sum_{i=1}^I\sum_{j=1}^J\sum_{k=1}^K\sum_{p=1}^Pu_{ip}(x_{ijk} - m_{pjk})^2 + \lambda_\mathbf{U}   {p}(\mathbf{U}), 
    \end{equation}
    \noindent
    where $m_{pjk}$ is the generic element of the array $\underline{\mathbf{M}}$ and $\lambda_{\mathbf{U}}$ calibrates the penalty's weight.
    The minimization of $f_1$ is carried out via an ALS procedure. Given $\mathbf{U}$, the optimal $\underline{\mathbf{M}}$ is obtained in closed form as the weighted average $m_{pjk} = \frac{\sum_{i=1}^I u_{ip} x_{ijk}}{\sum_{i=1}^I u_{ip}}$. Given $\underline{\mathbf{M}}$, the membership matrix $\mathbf{U}$ is updated using the entropic update rule $u_{ip} = \frac{e^{-d_{ip}^2/\lambda_\mathbf{U}}}{\sum_{h=1}^P e^{-d_{ih}^2/\lambda_\mathbf{U}}}$, \\ where $d_{ip}^2 = \sum_{j=1}^J\sum_{k=1}^K (x_{ijk} - m_{pjk})^2$. These two steps are alternated until convergence.
    \vspace{0.2cm}
    \item[b.] \textbf{Partitioning of variables} via fuzzy entropic $k$-means algorithm applied to $\underline{\mathbf{M}}$ by minimizing
    \begin{equation}\label{2}
        f_2(\mathbf{V},\underline{\mathbf{C}}) = \sum_{j=1}^J\sum_{k=1}^K\sum_{p=1}^P\sum_{q=1}^Qv_{jq}u_{+p}(m_{pjk} - c_{pqk})^2 + \lambda_\mathbf{V}   {p}(\mathbf{V}),
    \end{equation}
    \noindent
   where $u_{+p} = \sum_{i=1}^I u_{ip}$, $c_{pqk}$ is the generic element of $\underline{\mathbf{C}}$ and $\lambda_{\mathbf{V}}$ calibrates the penalty's weight. Also in this case, $f_2$ is minimized via ALS.
   \vspace{0.2cm}
    \item[c.] \textbf{Partitioning of occasions} via fuzzy entropic $k$-means algorithm applied to $\underline{\mathbf{C}}$ by minimizing
    \begin{equation}\label{3}
        f_3(\mathbf{W},\underline{\mathbf{G}}) = \sum_{k=1}^K\sum_{p=1}^P\sum_{q=1}^Q\sum_{r=1}^Rw_{kr}u_{+p}v_{+q}(c_{pqk} - g_{pqr})^2 + \lambda_\mathbf{W}   {p}(\mathbf{W}),
    \end{equation}
    \noindent
    where $v_{+q} = \sum_{j=1}^J v_{jq}$, $g_{pqr}$ is the generic element of $\underline{\mathbf{G}}$ and $\lambda_{\mathbf{W}}$ calibrates the penalty's weight. Similarly, $f_3$ is minimized via ALS.
\end{enumerate}

It is important to note that in this section the above procedures have been described using fuzzy entropic $k$-means as the fundamental clustering procedure, but other procedures can similarly be obtained by using different techniques.

Tandem procedures are easy to implement but have well-known limitations \citep{arabie1990three}. In the first step, the partition of objects is chosen independently of the partitions of variables and occasions, whereas in later steps the partitions are conditioned on previously estimated centroids. This asymmetry is not generally justified and can lead to blocks that are not globally optimal in a three-way sense. Moreover, the six possible tandem orders typically yield different solutions, highlighting order dependence. These limitations motivated the simultaneous approach we have introduced in section ($\ref{sec:simultaneous})$.



\subsubsection{Relationship between tandem and simultaneous procedures}\label{sec3.3}

In the tandem approach, fuzzy entropic $k$-means is applied sequentially to each way of the data. It is possible to show that the centroids along each way are updated step by step in the following way.
\begin{enumerate}
    \item Object centroids. Given fixed memberships $u_{ip}$, the object centroids $m_{pjk}$ that minimize (\ref{1}) are computed as weighted averages along the objects
    \begin{equation}
    m_{pjk} = \frac{\sum_{i=1}^I u_{ip} x_{ijk}}{\sum_{i=1}^I u_{ip}}. \label{10}
    \end{equation}
    \item Object-variable centroids. Using the object centroids $m_{pjk}$ estimated by the previous model and memberships $v_{jq}$, the variable centroids $c_{pqk}$ that minimize (\ref{2}) are updated as
    \begin{equation}
    c_{pqk} = \frac{\sum_{j=1}^J v_{jq} m_{pjk}}{\sum_{j=1}^J v_{jq}}. \label{11}
    \end{equation}
    \item Object-variable-occasion centroids. Finally, the object-variable-occasion centroids $g_{pqr}$ that minimize (\ref{3}) are computed using the object-variable centroids $c_{pqk}$ and memberships $w_{kr}$
    \begin{equation}
    g_{pqr} = \frac{\sum_{k=1}^K w_{kr} c_{pqk}}{\sum_{k=1}^K w_{kr}}. \label{12}
    \end{equation}
\end{enumerate}
\noindent
By substituting (\ref{10}) into (\ref{11}), we obtain
\begin{equation}
    c_{pqk} = \frac{\sum_{j=1}^J v_{jq} \left( \sum_{i=1}^I u_{ip} x_{ijk} / \sum_{i=1}^I u_{ip} \right)}{\sum_{j=1}^J v_{jq}}
= \frac{\sum_{i=1}^I \sum_{j=1}^J u_{ip} v_{jq} x_{ijk}}{\sum_{i=1}^I u_{ip} \sum_{j=1}^J v_{jq}}.\label{second}
\end{equation}
\noindent
Substituting (\ref{second}) into (\ref{12}) gives
\begin{align}\label{eq_tandem}
    & g_{pqr} = \frac{\sum_{k=1}^K w_{kr} \left( \sum_{i=1}^I \sum_{j=1}^J u_{ip} v_{jq} x_{ijk} / \sum_{i=1}^I u_{ip} \sum_{j=1}^J v_{jq} \right)}{\sum_{k=1}^K w_{kr}} = \\ \nonumber
    & \frac{\sum_{i=1}^I \sum_{j=1}^J \sum_{k=1}^K u_{ip} v_{jq} w_{kr} x_{ijk}}{\sum_{i=1}^I u_{ip} \sum_{j=1}^J v_{jq} \sum_{k=1}^K w_{kr}}.
\end{align}
\noindent
It has to be noted that (\ref{eq_tandem}) has the exact same form as the centroids in (\ref{obj_fe3km}). In fact, the minimum of (\ref{obj_fe3km}) with respect to $\underline{\mathbf{G}}$ is obtained by equating the partial derivative of $f_{FE3KM}$ with respect to each element $g_{pqr}$ to zero
\begin{align}
& \frac{\partial f_{FE3KM}}{\partial g_{pqr}} 
= \frac{\partial}{\partial g_{pqr}} \sum_{i=1}^{I} \sum_{j=1}^{J} \sum_{k=1}^{K} \sum_{p=1}^{P} \sum_{q=1}^{Q} \sum_{r=1}^{R} u_{ip} v_{jq} w_{kr} \left( x_{ijk} - g_{pqr} \right)^2 = \nonumber \\
& -2 \sum_{i=1}^{I} \sum_{j=1}^{J} \sum_{k=1}^{K} u_{ip} v_{jq} w_{kr} \left( x_{ijk} - g_{pqr} \right) = 0. \label{partialG}
\end{align}
Solving for $g_{pqr}$ gives the update rule, in terms of summations, for each $p,q,r$
\begin{equation}
g_{pqr} = \frac{\sum_{i=1}^{I} \sum_{j=1}^{J} \sum_{k=1}^{K} u_{ip} v_{jq} w_{kr} x_{ijk}}{\sum_{i=1}^{I}u_{ip}\sum_{j=1}^{J}v_{jq}\sum_{k=1}^{K}w_{kr}}. \label{Gsum}
\end{equation}
The partial second derivative, $\partial^2f_{FE3KM}/\partial g^2_{pqr} = 2u_{+p}v_{+q}w_{+r}$, is positive when clusters are non-empty ensuring that (\ref{Gsum}) is a minimum. This expression shows that each element $g_{pqr}$ of the core tensor is computed as a weighted average of the data entries $x_{ijk}$, with weights given by the membership degrees $u_{ip}$, $v_{jq}$, and $w_{kr}$ along the three ways. The expression in (\ref{gsumMatrix}) is the matrix representation of the same calculation.
\begin{equation}\label{gsumMatrix}
    \mathbf{G}_{P,QR} = (diag(\mathbf{U}^\prime\mathbf{1}_I))^{-1}\mathbf{U}^\prime\mathbf{X}_{I,JK}[\mathbf{W}(diag(\mathbf{W}^\prime\mathbf{1}_K))^{-1} \otimes \mathbf{V}(diag(\mathbf{V}^\prime\mathbf{1_J}))^{-1}]. 
\end{equation}
%
This shows that, at the level of centroid formulas, the sequential and simultaneous procedures are consistent in form. However, the optimization problems they solve are different. In the tandem approach, each way is updated only once, conditionally on centroids and memberships obtained in previous steps, and each step minimizes its own loss (e.g., \eqref{1}-\eqref{3}) subject to those conditions. In contrast, FE3KM repeatedly updates all modes and the core so as to (approximately) minimize the joint loss \eqref{obj_fe3km}. As a result, the tandem and simultaneous procedures generally converge to different solutions, with the latter better aligned to a global three-way LS criterion.
To further clarify the relationship between the simultaneous and tandem approaches, it can be shown that each objective function of the tandem procedures can be derived as a constrained version of the simultaneous objective function in (\ref{obj_fe3km}). By setting $\mathbf{V} = \mathbf{I}_J$ and $\mathbf{W} = \mathbf{I}_K$, the minimization of (\ref{obj_fe3km}) becomes (\ref{1}). As a second step, given the estimated optimal solutions $\hat{\mathbf{U}}$ and $\hat{\mathbf{\underline{M}}}$, we need to find the optimal expression for $\mathbf{V}$ and $\mathbf{\underline{C}}$ by minimizing 
\begin{equation}
    f(\mathbf{V}, \mathbf{\underline{C}}) 
= \sum_{j k}\sum_{pq} \hat{u}_{+p}v_{jq}(\hat{m}_{pjk} - c_{pqk})^2 + \lambda_\mathbf{V}  p(\mathbf{V}),
\end{equation}
\noindent
which is exactly (\ref{2}), given the optimal solution identified minimizing (\ref{1}). Finally, given $\hat{\mathbf{U}}$, $\hat{\mathbf{\underline{M}}}$, $\hat{\mathbf{V}}$ and $\hat{\mathbf{\underline{C}}}$, the minimization with respect to $\mathbf{W}$ and $\mathbf{\underline{G}}$ is given by
\begin{equation}
 f(\mathbf{W}, \mathbf{\underline{G}}) = \sum_{kpqr} \hat{u}_{+p}\hat{v}_{+q}w_{kr}(\hat{c}_{pqk} - g_{pqr})^2 + \lambda_\mathbf{W}  p(\mathbf{W}),
\end{equation}
\noindent
which corresponds to (\ref{3}). This decomposition highlights that the simultaneous procedure can be rewritten as the sum of the three objective functions of the tandem procedures. The key advantage of the proposed simultaneous method is that it jointly optimizes all three components at once, rather than performing three sequential optimizations that depend on the results of previous steps. Consequently, the simultaneous fuzzy entropic $k$-means ensures a more coherent and globally optimal estimation across all ways.
Another interesting result is the following. The previously described tandem procedure is the result of the sequential minimization of the three losses (\ref{1}), (\ref{2}) and (\ref{3}).
\vspace{0.3cm}

\noindent
\textbf{Proposition 2}: Conditional decomposition into sequential losses. For fixed row-stochastic $\mathbf{U}, \mathbf{V},$ and $\mathbf{W}$ with non-empty clusters, and intermediate means defined by ($\ref{10}- \ref{12} $). A simultaneous version could be obtained by minimizing the sum of the three equations $(\ref{1}), (\ref{2}), (\ref{3})$, that is
$f_{FE3KM}(\mathbf{U},\mathbf{V},\mathbf{W},\underline{\mathbf{G}}) = f_1(\mathbf{U},\underline{\mathbf{M}}) + f_2(\mathbf{V},\underline{\mathbf{C}}) + f_3(\mathbf{W},\underline{\mathbf{G}}).$

\vspace{0.2cm}
\noindent
\textbf{Proof}.
 This can be seen by observing that the three-way loss in (\ref{obj_fe3km}) can be decomposed as
\begin{align}\label{d1}
    & \nonumber    \sum_{ijkpqr}u_{ip}v_{jq}w_{kr}(x_{ijk} - c_{pqk} + c_{pqk} - g_{pqr})^2 + \lambda_\mathbf{U} {p}(\mathbf{U}) + \lambda_\mathbf{V}   {p}(\mathbf{V}) + \lambda_\mathbf{W}   {p}(\mathbf{W}) = \\ & 
    \nonumber \sum_{ijkpq}u_{ip}v_{jq}(x_{ijk} - c_{pqk})^2 + \sum_{kpqr}u_{+p}v_{+q}w_{kr}(c_{pqk} - g_{pqr})^2\ + \\
    & \lambda_\mathbf{U}   {p}(\mathbf{U}) + \lambda_\mathbf{V}   {p}(\mathbf{V}) + \lambda_\mathbf{W}   {p}(\mathbf{W}),
\end{align}
\noindent
since
\begin{align}
    & \nonumber    \sum_{ijkpqr} u_{ip}v_{jq}w_{kr}(x_{ijk}-c_{pqk})(c_{pqk}-g_{pqr}) = \\ 
    & \sum_{kpqr}w_{kr}(c_{pqk}-g_{pqr})\sum_{ij}u_{ip}v_{jq}(x_{ijk} - c_{pqk}) = 0,
\end{align}
\noindent
where $\sum_{ij}u_{ip}v_{jq}(x_{ijk} - c_{pqk}) =$ 0, because $c_{pqk}$ is the appropriate weighted mean of the observations. The first term in the last expression of (\ref{d1}) can be further decomposed as
\begin{align}\label{d2}
    & \sum_{ijkpq}u_{ip}v_{jq}(x_{ijk} - m_{pjk} + m_{pjk} - c_{pqk})^2 = \nonumber \\
    & \sum_{ijkp}u_{ip}(x_{ijk} - m_{pjk})^2 +\sum_{jkpq}u_{+p}v_{jq}(m_{pjk} - c_{pqk})^2,
\end{align}
\noindent
since
\begin{align}
    & \nonumber    \sum_{ijkpq} u_{ip}v_{jq}(x_{ijk}-m_{pjk})(m_{pjk}-c_{pqk}) = \\     
    & \sum_{jkpq}v_{jq}(m_{pjk}-c_{pqk})\sum_{i}u_{ip}(x_{ijk} - m_{pjk}) = 0.
\end{align}
\noindent
Substituting (\ref{d2}) in (\ref{d1}), we get the decomposition of (\ref{obj_fe3km}) in 
\begin{align}\label{d3}
\nonumber
    & \sum_{ijkp}u_{ip}(x_{ijk} - m_{pjk})^2 +\sum_{jkpq}u_{+p}v_{jq}(m_{pjk} - c_{pqk})^2\ + \\ \nonumber
    & \sum_{kpqr}w_{kr}u_{+p}v_{+q}(c_{pqk} - g_{pqr})^2 + \lambda_\mathbf{U}   {p}(\mathbf{U}) + \lambda_\mathbf{V}   {p}(\mathbf{V}) + \lambda_\mathbf{W}   {p}(\mathbf{W}) = \\ 
    & (\ref{1}) + (\ref{2}) + (\ref{3}).
\end{align}

This identity does not imply that independently minimizing the three tandem stages minimizes the joint objective function. Indeed, this solution represents an initial estimation of the objective function $(\ref{obj_fe3km})$.
By contrast, simultaneous estimation repeatedly updates all three partitions as we will see in the next section.

\subsubsection{Independent procedures}\label{sec:independent}

This section illustrates an even simpler way to obtain a fuzzy partition of the three ways is to cluster each of them separately, without any reference to the other two. Specifically, fuzzy entropic $k$-means can be applied to the three matricized arrays $\mathbf{X}_{I,JK}$,
$\mathbf{X}_{J,IK}$ and $\mathbf{X}_{K,IJ}$, which yields the three row-stochastic membership matrices $\mathbf{U}$, $\mathbf{V}$ and $\mathbf{W}$. For the objects, for instance, $\mathbf{U}$ minimizes $\sum_{ijkp} u_{ip}(x_{ijk}-m_{pjk})^2+\lambda_{\mathbf{U}}\,p(\mathbf{U})$. The membership updates are the softmax rules of
Step 2 in the non-accelerated algorithm in Section~\ref{sec3.4}, and the same applies, with the proper matricization, to $\mathbf{V}$ and $\mathbf{W}$. Once the three partitions are available, the block centroids are obtained by minimizing, with respect to the array $\underline{\mathbf{G}}=[g_{pqr}]$ only, the loss
\begin{equation}\label{eq:indep}
f(\underline{\mathbf{G}})=\sum_{i=1}^{I}\sum_{j=1}^{J}\sum_{k=1}^{K}\sum_{p=1}^{P}\sum_{q=1}^{Q}\sum_{r=1}^{R}
u_{ip}\,v_{jq}\,w_{kr}\,\bigl(x_{ijk}-g_{pqr}\bigr)^2 ,
\end{equation}
which is Equation~\eqref{eq_3.2} (the unregularized loss) evaluated at fixed $(\mathbf{U},\mathbf{V},\mathbf{W})$. Since $\mathbf{U}$, $\mathbf{V}$ and $\mathbf{W}$ are held fixed, the penalty terms are constant and the solution is the weighted mean in Equation~(\ref{Gsum}).

This procedure is easy to implement, fully parallelizable across modes, and not affected by the order
in which the modes are treated, unlike the tandem procedures. However, it has two main drawbacks.
First, the minimization is suboptimal with respect to the three-way criterion: the loss
\eqref{eq:indep} is minimized only with respect to $\underline{\mathbf{G}}$, while the memberships
are never chosen to reduce it. Each of them is determined by the distances in a single unfolded space, in which the
heterogeneity coming from the other two ways is not summarized by the partitions of those ways.
Second, the three classifications are obtained independently of each other, so cross-way interactions are ignored, and blocks that are coherent along each way separately need not be homogeneous as three-way blocks. The tandem procedures partially remove these limitations by conditioning each step on the previous ones, and FE3KM removes them by optimizing
$(\mathbf{U},\mathbf{V},\mathbf{W},\underline{\mathbf{G}})$ jointly. By Proposition 3, FE3KM started from the independent solution cannot increase the objective value, so the independent partitions are a natural and inexpensive ``rational'' initialization for the simultaneous algorithm.

\subsection{Algorithms for fuzzy entropic triple \texorpdfstring{\textit{k}}--means}\label{sec3.4}

The most straightforward algorithm to minimize (\ref{obj_fe3km}) consists in updating the core $\mathbf{G}_{P,QR}$ using (\ref{gsumMatrix}) and then alternatively updating one partition at a time, given the others. This is a block-coordinate estimation 
For fixed cluster numbers and regularization coefficients, Eq. ($\ref{eq_3.2}$) is minimized by  updating sequentially blocks of parameters $\mathbf{\underline{G}}, \mathbf U, \mathbf V,$ and $\mathbf W$, where each update minimizes the objective ($\ref{eq_3.2}$)  with respect to one block while holding the remaining blocks fixed.

\newpage

\noindent\rule{\textwidth}{0.4pt}
\vspace{-0.6cm}
\begin{center}\textbf{Non accelerated algorithm}\end{center}
\vspace{-0.4cm}
\rule{\textwidth}{0.4pt} 
\noindent
\noindent
\textbf{Initialization}. Initial values are chosen for $\mathbf{U}$, $\mathbf{V}$, and $\mathbf{W}$. Such values can be chosen randomly or in a rational way (e.g., based on independent partitioning).

\noindent
\textbf{Step 1}. \textbf{Update} $\mathbf{G}_{P,QR}$ according to (\ref{gsumMatrix}).

\noindent
\textbf{Step 2}. \textbf{Update} $\mathbf{U}$. Define

\begin{equation*}
    d_{ip}^2 = \sum_{jkqr}v_{jq}w_{kr}(x_{ijk}-g_{pqr})^2,\ \forall i=1,\ldots,I, p = 1,\ldots, P,
\end{equation*}

\noindent

Then, $u_{ip} = \frac{e^{-d_{ip}^2/\lambda_\mathbf{U}}}{\sum_{h=1}^Pe^{-d_{ih}^2/\lambda_\mathbf{U}}},\ \forall i=1,\ldots,I,p=1,\ldots,P$.

\vspace{0.2cm}

\noindent
\textbf{Step 3}. \textbf{Update} $\mathbf{V}$. Define

\begin{equation*}
    d_{jq}^2 = \sum_{ikpr}u_{ip}w_{kr}(x_{ijk}-g_{pqr})^2,\ \forall j=1,\ldots,J,q = 1,\ldots, Q,
\end{equation*}

\noindent
then $v_{jq} = \frac{e^{-d_{jq}^2/\lambda_\mathbf{V}}}{\sum_{h=1}^Qe^{-d_{jh}^2/\lambda_\mathbf{V}}},\ \forall j=1,\ldots,J,q=1,\ldots,Q$.

\vspace{0.2cm}

\noindent
\textbf{Step 4}. \textbf{Update} $\mathbf{W}$. Define

\begin{equation*}
    d_{kr}^2 = \sum_{ijpq}u_{ip}v_{jq}(x_{ijk}-g_{pqr})^2,\ \forall k=1,\ldots,K,r = 1,\ldots, R,
\end{equation*}

\noindent
then $w_{kr} = \frac{e^{-d_{kr}^2/\lambda_\mathbf{W}}}{\sum_{h=1}^Re^{-d_{kh}^2/\lambda_\mathbf{W}}},\ \forall k=1,\ldots,K,r=1,\ldots,R$.

\vspace{0.2cm}

\noindent
\textbf{Stopping rule}. The function value in (\ref{obj_fe3km}) is computed for the current values of the parameters. If such updated values have decreased considerably (more than an arbitrarily small value), the parameter matrices are updated once more according to Steps 1--4. Otherwise, the process is considered to have converged.

\noindent\rule{\textwidth}{0.4pt}


When one or more ways become very large (e.g., $\geq 100$), the assignment step can be accelerated by computing the distances on the clustered centroids instead of on the full data. No dimension reduction of the data or of the model is involved, and the memberships
obtained are the same as in the non-accelerated algorithm. As an example, the update of $\mathbf{W}$ could be performed considering (\ref{d1}), where the dimension of the data would be $P \times Q \times K$ instead of $I \times J \times K$. Equivalent derivations can be obtained for the other modes.

The accelerated algorithm can then be formulated as follows.

\noindent\rule{\textwidth}{0.4pt}
\vspace{-0.6cm}
\begin{center}\textbf{Accelerated algorithm}\end{center}
\vspace{-0.4cm}
\rule{\textwidth}{0.4pt} 
\noindent
\textbf{Initialization}. Initial values are chosen for $\mathbf{U}$, $\mathbf{V}$, and $\mathbf{W}$. Such values can be chosen randomly or in a rational way (e.g., based on independent partitioning).

\noindent
\textbf{Step 1}. \textbf{Update} $\mathbf{G}_{P,QR}$ according to (\ref{gsumMatrix}).

\vspace{0.2cm}
\noindent
\textbf{Step 2}. \textbf{Update} $\mathbf{U}$. Define distances in the clustered $\mathcal{R}^{I \times Q \times R}$ space as

\begin{equation*}
    \tilde{d}_{ip}^2 = \sum_{qr}v_{+q}w_{+r}(c'_{iqr}-g_{pqr})^2,\ \forall i=1,\ldots,I, p = 1,\ldots, P.
\end{equation*}

\noindent
Since  $d_{ip}^2 = C_i + \tilde{d}_{ip}^2$, where $C_i = \sum_{jkqr} v_{jq}w_{kr}(x_{ijk}-c^\prime_{iqr})^2$, and
 $c^\prime_{iqr} = \frac{\sum_{j=1}^J\sum_{k=1}^Kv_{jq}w_{kr}x_{ijk}}{\sum_{j=1}^J\sum_{k=1}^Kv_{jq}w_{kr}},$ 

\vspace{0.2cm}

\noindent
we have that $C_i$ is independent of $p$. Thus, it is easy to show that $u_{ip}=\frac{e^{-(C_i + \tilde{d}_{ip}^2)/\lambda_\mathbf{U}}}{\sum_{h=1}^Pe^{-(C_i + \tilde{d}_{ih}^2)/\lambda_\mathbf{U}}}= \frac{e^{-\tilde{d}_{ip}^2/\lambda_\mathbf{U}}}{\sum_{h=1}^Pe^{-\tilde{d}_{ih}^2/\lambda_\mathbf{U}}}$, since the factor $e^{-(C_i/\lambda_\mathbf{U})}$ in the numerator and denominator of the first equation cancels. 
Then, memberships are updated with this simplified formula.

Analogous simplifications can be stated for the other modes and their proof is omitted.

\vspace{0.2cm}
\noindent
\textbf{Step 3}. \textbf{Update} $\mathbf{V}$. Define distances in the clustered $\mathcal{R}^{P \times J \times R}$ space as

\begin{equation*}
    \tilde{d}_{jq}^2 = \sum_{pr}u_{+p}w_{+r}(c''_{pjr}-g_{pqr})^2,\ \forall j=1,\ldots,J,q = 1,\ldots, Q,
\end{equation*}

\noindent
where $c''_{pjr} = \frac{\sum_{i=1}^I\sum_{k=1}^Ku_{ip}w_{kr}x_{ijk}}{\sum_{i=1}^I\sum_{k=1}^Ku_{ip}w_{kr}}, p=1,\ldots,P,j=1,\ldots,J,r=1,\ldots,R$. Then, memberships are updated as in Step 3 of the non-accelerated algorithm, with the simplified formula.

\vspace{0.2cm}
\noindent
\textbf{Step 4}. \textbf{Update} $\mathbf{W}$. Define distances in the clustered $\mathcal{R}^{P \times Q \times K}$ space as

\begin{equation*}
    \tilde{d}_{kr}^2 = \sum_{pq}u_{+p}v_{+q}(c_{pqk}-g_{pqr})^2,\ \forall k=1,\ldots,K,r = 1,\ldots, R,
\end{equation*}

\noindent
then, memberships are updated as in Step 4 of the non-accelerated algorithm, with the simplified formula.

\vspace{0.2cm}
\noindent
\textbf{Stopping rule}. The function value in (\ref{d3}) is computed for the current values of the parameters. If such updated values have decreased considerably (more than an arbitrarily small value), the parameter matrices are updated once more according to Steps 1--4. Otherwise, the process is considered to have converged.

\noindent\rule{\textwidth}{0.4pt}

The procedure is run from multiple initializations (multi-start approach). For fixed cluster numbers and regularization coefficients, the solution with the smallest final objective is retained.
\vspace{0.3cm}
\newline \noindent
\textbf{Proposition 3}: Monotonicity of the objective function ($\ref{obj_fe3km}$) sequences.
For fixed positive regularization coefficients, positive initial cluster masses (i.e., no cluster is empty), and exact sequential block updates, both algorithms generate a non increasing sequence of objective values. This sequence converges to a finite value.
\newline
\noindent
\textbf{Proof}. Each centroid update and each membership update is a conditional minimization of the same objective. Therefore, no update increases ($\ref{obj_fe3km}$). Moreover, $L \geq 0$, $p(\mathbf{U}) \geq - I \log P$, $p(\mathbf{V}) \geq -J log Q$, and $p(\mathbf{W}) \geq -K \log R$. Hence: 
$f_{FE3KM}(\mathbf{U},\mathbf{V},\mathbf{W},\underline{\mathbf{G}}) \geq -\lambda_\mathbf U I \log P -\lambda_\mathbf V J \log Q -\lambda_\mathbf W K \log R.$
Therefore, the objective sequence is monotone and bounded below. The lower bound is finite and is strictly negative when the coefficients $\lambda$ are positive and at least one mode has more than one cluster. The objective function (\ref{obj_fe3km}) itself can be negative, zero, or positive.

\vspace{0.3cm}
This result establishes convergence of objective values, not global optimality or, by itself, convergence of the parameter sequence. Because the joint problem is nonconvex, different initializations can produce different fitted solutions. While the update steps do not guarantee global optimality, running the algorithm multiple times with different random initializations increases the likelihood of reaching the global minimum. In our simulation study and application, 30 random initializations have been considered.

In practice, the algorithm is terminated when the relative decrease in the objective function falls below a predefined tolerance. 


In our implementation, we adopt a random fuzzy initialization strategy: each membership matrix is generated by drawing its entries from independent uniform distributions on $[0,1]$ and then normalizing each row to sum to one. Specifically, for the object membership matrix $\mathbf{U}$, we generate an $I \times P$ matrix of independent $\texttt{Unif}(0,1)$ random variates and apply row-wise normalization such that $\sum_{p=1}^P u_{ip} = 1$ for all $i$. The same procedure is applied to initialize $\mathbf{V}$ and $\mathbf{W}$. This approach ensures that the initial partitions are fuzzy, with each entity belonging to all clusters with positive membership values, while respecting the stochastic constraints on the rows.

\vspace{0.2cm}

\textbf{Cross-validation and tuning-parameter selection}. For fixed cluster numbers $(P, Q, R)$, the tuning parameters 
$\lambda_\mathbf{U}$, $\lambda_\mathbf{V}$, $\lambda_\mathbf{W}$ are selected via 5-fold cross-validation. The observed entries of the array are divided into five approximately equal validation sets, ensuring that each object, variable, and occasion remains represented in the training entries of every fold. This procedure assesses prediction of unobserved entries for entities represented in the training data. The same folds are used for all candidate tuning parameters and cluster configurations. Any data-dependent centering or scaling is estimated from the training entries only and applied unchanged to the validation entries. For each candidate $\lambda$ triplet and each fold, FE3KM is fitted using only the training entries. The weighted-distance sums in the objective and membership updates are restricted to these entries, while the entropy terms retain their original form. The centroid update is correspondingly replaced by
$g_{pqr}$ = [sum over training entries $(i, j, k)$ of $u_{ip}v_{jq}w_{kr}x_{ijk}]$ / [sum over training entries $(i, j, k)$ of $u_{ip}v_{iq}w_{kr}$]. Each fit uses multiple initializations, retaining the solution with the smallest training objective for the given $\lambda$ triplet.

Thus, For each candidate regularization triplet and each training fold, all membership matrices $\mathbf{U}, \mathbf{V},$ and $\mathbf{W}$ and the core $\underline{\mathbf{G}}$ are jointly re-estimated by alternating minimization using only the training entries. Thus, the fitted memberships and centroids depend on both the regularization triplet and the training fold. These estimates are held fixed only when predicting and scoring the corresponding validation entries. After selecting the triplet that minimizes the pooled validation prediction error, all membership matrices and the core are re-estimated on the complete array.
During the algorithm’s iterations, extremely small membership values may be represented as zero because computers store numbers with limited precision. Small but positive cluster masses do not, by themselves, require reassignment. If a cluster mass becomes numerically zero and stable calculations cannot resolve the problem, a corrective reassignment may be proposed that preserves the row-stochastic constraints, followed by recomputation of the core G. The correction is accepted only if the full objective function $(\ref{obj_fe3km})$, including the entropy terms, does not increase. Otherwise, the correction is rejected or the run is restarted from a new initialization.

\subsection{Sum-of-squares decomposition and descriptive interpretation}\label{devdec}

In Subsection \ref{sec3.3}, we have seen how the fit can be decomposed in the sequential contribution of each way to the total within. In this subsection, we will show how the total deviance can be decomposed into the sum of the within and the between deviance. This assures that our method tends to group similar entities into the same cluster (within minimization) but it also tends to classify into distinct clusters different entities. Suppose that the data array $\underline{\mathbf{X}}$ is centered with respect to the grand-mean. 

\vspace{0.3cm}

\noindent
\textbf{Proposition 4}: The total deviance, i.e., the sum of squares $TSS=\|\underline{\mathbf X}\|^2=\sum_{ijk}x_{ijk}^2$ of $\underline{\mathbf X}$, can be decomposed into within and between components as follows $TSS = WSS + BSS$, where
\begin{align}\label{dec2} 
\newline
TSS = \|x_{ijk}\|^2 = \sum_{ijk} x_{ijk}^2 =   \sum_{ijkpqr}u_{ip} v_{jq} w_{kr} x_{ijk}^2 \\ 
WSS  = \sum_{ijkpqr} u_{ip} v_{jq} w_{kr} \left( x_{ijk} - g_{pqr} \right)^2\ \\
BSS = \sum_{pqr} u_{+p} v_{+q} w_{+r} g_{pqr}^2,
\end{align}
\noindent
and where 
\begin{equation}\label{centroids}
   g_{pqr} = \sum_{ijk} u_{ip} v_{jq} w_{kr} x_{ijk} \left(u_{+p}v_{+q} w_{+r} \right)^{-1}, 
\end{equation}
\noindent
\textbf{Proof}. The identity is immediately proved if  the cross-term between $WSS$ and $BSS$ vanishes, leaving $WSS$ and $BSS$. Indeed,

\vspace{-0.8cm}

\begin{align}
& \sum_{ijkpqr}u_{ip} v_{jq} w_{kr} (x_{ijk} - g_{pqr}) g_{pqr} = \sum_{pqr}\left( \sum_{ijk} u_{ip} v_{jq} w_{kr} x_{ijk} g_{pqr} - \sum_{ijk} u_{ip} v_{jq} w_{kr} g_{pqr}^2 \right) = \nonumber \\
& \sum_{pqr} \left[ \sum_{ijk} u_{ip}v_{jq} w_{kr} x_{ijk} \right] g_{pqr} - \sum_{pqr} u_{+p} v_{+q} w_{+r} g_{pqr}^2 \nonumber \overset{(\ref{centroids})}= \\
& \sum_{pqr}  \left[u_{+p} v_{+q} w_{+r} g_{pqr}\right]g_{pqr} - \sum_{pqr}  u_{+p} v_{+q} w_{+r} g_{pqr}^2 = 0.
\end{align}

\noindent

The identity holds whenever the core is updated to the weighted means of the current memberships, and it is valid for any partition $\mathbf{U}, \mathbf{V}$ and $\mathbf{W}$. Grand-mean centering simplifies the expression, but is not otherwise necessary.

The decomposition follows from normalized weights and weighted-mean centroids, not from entropy regularization itself. In conventional exponent-m weighted-distance formulations, the corresponding cross-term also vanishes when the centroids use the same exponentiated weights. However, those weights generally no longer sum to one across blocks for each observation. The resulting decomposition concerns a membership-weighted total rather than the ordinary $TSS$. After normalizing exponentiated weights, the centroids must also be recomputed using those normalized weights for the new decomposition to hold.

\vspace{0.3cm}

\noindent
\textbf{Remark 2}: The same type of decomposition cannot be found by using the  exponent-$m$ criterion (used in fuzzy $c$-means approach). In fact, suppose an alternative method uses weights: $b_{ijk,pqr} = u^m_{ip} v^m_{jq} w^m_{kr}$, and computes centroids $h_{pqr}$ using those same weights. Then: $\sum_{ijk} b_{ ijk,pqr} (x_{ijk} - h_{pqr}) = 0$, thus cross-term still vanishes. However, the combined weights now sum to: $s_{ijk} = \sum_{pqr} b_{ijk,pqr} = (\sum_{p}  u^m_{ip})(\sum_{q}  v^m_{jq})(\sum_{r}  w^m_{kr}) < 1$, when $m>1$, and membership are fuzzy. Thus, $TSS_m =\sum_{ijk}  s_{ijk}  (x_{ ijk})^2 = WSS_m + BSS_m$, where $WSS_m$ and $BSS_m$ use the exponent-$m$ weights and their associated centroids. Thus $TSS_m$ is not ordinary $TSS$. To obtain $TSS$, new weights have to be defined: $ a_{ijk,pqr} = b_{ ijk,pqr} / s_{ijk}$, so that $\sum_{prq} a_{ijk,pqr}=1$. Thus, $TSS = WSS_a + BSS_a$. However,  this last is not the decomposition of the exponent-$m$ criterion originally considered.

\vspace{0.3cm}
When $TSS > 0$, $BSS/TSS$ summarizes the proportion of ordinary variation represented by between-block centroid differences. It is not generally the same as a reconstruction-based coefficient of determination, because equation $L(\mathbf{U},\mathbf{V},\mathbf{W},\underline{\mathbf{G}})$ in Proposition 1, distinguishes $WSS$ from the Tucker reconstruction error.

\vspace{0.3cm}

\noindent
\textbf{Remark 3}: Since $WSS$ = $L(\mathbf{U},\mathbf{V},\mathbf{W},\underline{\mathbf{G}})$ the objective function $(\ref{obj_fe3km})$ can be linked to the decomposition by
$f_{FE3KM}(\mathbf{U},\mathbf{V},\mathbf{W},\underline{\mathbf{G}}) = TSS - BSS + \lambda_\mathbf{U}   {p}(\mathbf{U}) + \lambda_\mathbf{V}   {p}(\mathbf{V}) + \lambda_\mathbf{W}   {p}(\mathbf{W})$.
\vspace{0.2cm}
\newline \noindent
Thus, for fixed data and regularization coefficients, minimizing $f_{FE3KM}(\mathbf{U},\mathbf{V},\mathbf{W},\underline{\mathbf{G}})$ is therefore equivalent to maximizing
$BSS - \lambda_\mathbf{U}   {p}(\mathbf{U}) - \lambda_\mathbf{V}   {p}(\mathbf{V}) - \lambda_\mathbf{W}   {p}(\mathbf{W}).$

\vspace{0.3cm}

\noindent
\textbf{Proposition 5} (Cluster contributions to between-block variation): Denote cluster masses by $u_{+p} = \sum_i u_{ip}, v_{+q} = \sum_j v_{jq}$ and $w_{+r} = \sum_kw_{kr}$, for $\mathbf{U},\mathbf{V}, \mathbf{W}$. The contributions of unit cluster $p$, variable cluster $q$, and occasion cluster $r$ to between-block variation are, respectively,

\vspace{0.2cm}
$B_p^{(\mathbf{U})}$ =  $u_{+p} \sum _{qr} v_{+q} w_{+r} g^2_{pqr}$, $B_q^{(\mathbf{V})}$ =  $v_{+q} \sum _{pr} u_{+p} w_{+r} g^2_{pqr}$, $B_r^{(\mathbf{W})}$ =  $w_{+r} \sum _{pq} u_{+p} v_{+q}  g^2_{pqr}$.

\vspace{0.2cm}
These contributions are nonnegative and satisfy $\sum_p B_p^{(\mathbf{U})} = \sum_q B_q^{(\mathbf{V})} = \sum_r B_r^{(\mathbf{W})} = BSS$.
\noindent
When $TSS > 0$, the ratios:  $B_p^{(\mathbf{U})}/TSS$;  $B_q^{(\mathbf{V})}/TSS$, and $B_r^{(\mathbf{W})}/TSS$ express the respective cluster contributions on the total-variation scale; within each mode, they sum to $BSS/TSS$. 
\newline
\noindent
When $BSS$ > 0, normalization by $BSS$, i.e.: $B_p^{(\mathbf{U})}/BSS$;  $B_q^{(\mathbf{V})}/BSS$, and $B_r^{(\mathbf{W})}/BSS$  instead gives each cluster's share of between-block variation, with shares summing to one within each mode.

\vspace{0.2cm}

\noindent
\textbf{Proposition 6} (Gain from clustering in explained deviance). Let us suppose that we want to quantify the gain in explained deviance when moving from a model with no clustering of the units, i.e. $P = 1$, to a partition into several clusters one ($P>1$) while holding the variable and occasion memberships \textbf{V} and \textbf{W} fixed. This mirrors the logic of classical ANOVA: we ask how much additional variability is explained by distinct groups of elements. Assume that the data are centered and that the block centroids are the weighted means in Equation (\ref{centroids}), with positive cluster masses. When elements are partitioned into clusters, the explained deviance is
\begin{equation}
    \sum_{pqr} u_{+p}v_{+q}w_{+r}g_{pqr}^2,
\end{equation}
\noindent
while when all units are treated as a single group the explained deviance is
\begin{equation}
    \sum_{qr} I  v_{+q}w_{+r}g_{\cdot qr}^2,
\end{equation}
\noindent
where $g_{\cdot qr}$ is the centroid pooled across all units for variable cluster $q$ and occasion cluster $r$, i.e.
\begin{align}
    & g_{\cdot qr} = \frac{1}{I} \sum_{ip}u_{ip}c'_{iqr}=\frac{1}{I}\sum_{i}u_{+p}g_{pqr}.  
\end{align}
\noindent
The conditional increase in explained deviance is therefore
\begin{equation}\label{decomp}   \Delta^{(\mathbf{U})}=\sum_{pqr} u_{+p}v_{+q}w_{+r}g_{pqr}^2 - \sum_{qr}I  v_{+q}w_{+r}g_{\cdot qr}^2 = \sum_{pqr}u_{+p}v_{+q}w_{+r}(g_{pqr}-g_{.qr})^2 \ge 0.
\end{equation}
\noindent
This follows by expanding the square and using: $\sum_p u_{+p} g_{pqr} =Ig_{.qr}$. The expression measures how much additional variability is captured, allowing different units clusters to have different centroids, conditional on the variables and occasion partitions. Knowing that $g_{pqr} = \frac{1}{u_{+p}}\sum_{i}u_{ip}c'_{iqr}$, (\ref{decomp}) can be rewritten as
\begin{equation}\label{dev}
     \Delta^{(\mathbf{U})}=\sum_{qr}v_{+q}w_{+r}\left[\sum_{p}u_{+p}g_{pqr}^2 - I\left( \frac{1}{\sum_pu_{+p}}\sum_{p}u_{+p}g_{pqr}\right)^2\right]. 
\end{equation}
\noindent
Thus, the gain is a weighted sum of between-unit-cluster deviances (sums of squares), evaluated across the variable-occasion cluster combinations. The weights account for the cluster masses in the other two modes. This provides a natural generalization of the ANOVA between-group sum of squares to the fuzzy three-way setting.
This contrast provides a descriptive measure of the contribution of the unit partition to explained variability. When $TSS > 0$, $\Delta^{(\mathbf{U})}/TSS$ expresses this gain as a proportion of total variation. Analogous contrasts can be obtained for variables and occasions by pooling the corresponding mode while holding the other two memberships fixed. These are conditional descriptive comparisons, not significance tests or comparisons between separately re-estimated models.


\subsection{Theoretical comparison with alternative three-way clustering methods}\label{compare}

FE3KM can be positioned with respect to the families of three-way clustering methods introduced in Section \ref{sec:introduction}, which are grouped according to what is clustered and how. We compare them along four dimensions: ($i$) the objective function, ($ii$) the type of memberships and constraints, ($iii$) the way in which the three modes are clustered (simultaneously or sequentially), and ($iv$) whether the ordinary Total Sum of Squares ($TSS$) admits an exact within/between decomposition. Table~\ref{tab:comparison} summarizes the comparison.

\paragraph{Least-squares partitioning of all three modes.}
\noindent
\citet{rocci2003threemode}, \citet{kiers2004} and \citet{schepers06} partition objects, variables and occasions and approximate the array by block centroids. \citet{kiers2004} studies the non-overlapping model in which all three ways are partitioned and the data are approximated by a Tucker-form reconstruction with binary memberships. For one-hot $\mathbf U$, $\mathbf V$, $\mathbf W$ the dispersion term of Proposition~1 vanishes, so (\ref{eq_3.2}) coincides with this reconstruction criterion, and the weighted-mean update of the core reduces to its ordinary LS form. FE3KM is therefore a fuzzy, entropy-regularized extension of this model and recovers it in the limit $\lambda\to0$, where memberships become hard. With respect to standard Tucker3 and related component models \citep{Tucker1966}, the row-stochastic nonnegative constraints on $\mathbf U$, $\mathbf V$, $\mathbf W$ turn component scores into memberships, so that the model is a pure clustering method and the core $\underline{\mathbf G}$ contains block centroids. The methods of this family are simple and scalable, but their hard assignments provide no description of the uncertainty of the classification; the entropy term of FE3KM supplies it, with a tunable level of fuzziness.

\begin{table}[ht!]
\centering
\caption{Conceptual comparison of three-way clustering approaches}
\label{tab:comparison}
\small
\begin{tabular}{p{4.3cm}p{2.2cm}p{2.6cm}p{2.6cm}}
\toprule
\textbf{Approach} & \textbf{Memberships} & \textbf{Formulation} & \textbf{Exact $TSS$ decomp.} \\
\midrule
LS partitioning of all three modes
  & hard & Least-Squares & yes (hard case) \\
  \hline
Component analysis + partition of the units
  & crisp or fuzzy (one mode) & Least-Squares & not a focus \\
  \hline
Factorization, then clustering
  & hard & two-step & not applicable \\
  \hline
Tensor block models, spectral and tri-clustering methods
  & hard & generative / spectral & not a focus \\
  \hline
Fuzzy memberships: fuzzy tri-clustering and co-clustering
  & fuzzy (all modes) & aggregation-type / entropy-regularized & not a focus \\
  \hline
Fuzzy memberships: mixed-membership blockmodel
  & mixed (simplex) & generative, spectral estimation & not a focus \\
  \hline
Probabilistic mixtures
  & posterior probabilities & likelihood & not applicable \\
  \hline
FE3KM
  & fuzzy (entropic, all modes) & Least-Squares & yes (any row-stochastic memberships) \\
\bottomrule
\end{tabular}
\end{table}

\paragraph{Component analysis combined with a partition of the units.}
\noindent
\citet{Rocci2005} and \citet{Vichi2007} combine a crisp or fuzzy partition of one way (the units) with component analysis of the other two ways. Since \citet{Rocci2005} already allow a fuzzy partition of one way, fuzzy memberships are not new in three-way analysis; what has rarely been done is to extend them to all ways. FE3KM does it, has no component matrices, and decouples the numbers of clusters $(P,Q,R)$ from the number of components. \citet{Schoonees2021} cluster the matrices of a three-way array through a LS decomposition into mean, row, column and interaction effects, which is a different clustering structure from the three-mode blocks of FE3KM.

\paragraph{Factorization followed by clustering.}
\noindent
Methods that first estimate a low-rank CANDECOMP/PARAFAC or Tucker factorization and then apply $k$-means or spectral clustering in the factor space \citep{huang08,sun16,wu16} optimize one criterion for the factorization and another for the clustering. FE3KM instead optimizes a single integrated LS criterion (Equation~(\ref{eq_3.2})) in which clustering and factorization are performed simultaneously. The tandem procedures used as benchmarks in our simulation study belong to this family.

\paragraph{Tensor block models, spectral and tri-clustering methods.}
\noindent
Tensor block models, tensor co-clustering and spectral tri-clustering \citep{wang21,feizi17,andriantsiory21,battaglia23} target latent block parameters and offer statistical guarantees under their assumptions, whereas FE3KM is a descriptive LS criterion whose guarantees concern the objective (Proposition~3) rather than estimation consistency.

\paragraph{Fuzzy memberships.}
\noindent
Fuzzy memberships have been introduced in three-way clustering in several ways \citep{Rocci2005,coppi10,liu15,honda21,Agterberg2025}. Fuzzy tri-clustering and three-mode fuzzy co-clustering \citep{liu15,honda21} assign fuzzy memberships to all three ways and, in \citet{liu15}, regularize them with one entropy term per way. They are, however, based on aggregation-type criteria rather than on a LS distance to a full array of block centroids with three separately indexed partitions. To our knowledge, they do not provide a decomposition of the ordinary $TSS$ into within- and between-block components, nor the way- and cluster-level contributions of Subsection~\ref{devdec}. The tensor mixed-membership blockmodel of \citet{Agterberg2025} is generative, with identifiability and estimation guarantees, while FE3KM is descriptive. In FE3KM the decomposition of Subsection~\ref{devdec} holds for any row-stochastic memberships with weighted-mean centroids and normalized product weights, so that contributions to explained deviance can be attributed to each way and to each cluster within a way. With exponent-$m$ weights, in contrast, the weights do not sum to one, and the decomposition concerns a weighted total rather than the ordinary $TSS$ (Remark~2).

\paragraph{Probabilistic mixtures.}
\noindent
\citet{Vermunt2007} proposes a hierarchical mixture model for three-way data, in which clustering is based on posterior probabilities and inference is likelihood-based. It targets latent mixing proportions under distributional assumptions, whereas FE3KM makes no distributional assumption.

This subsection showed that these families target different quantities (hard partitions, latent mixing proportions, co-occurrence structure), so their estimates are not directly comparable on a common metric without favoring one formulation. We therefore position the methods conceptually: FE3KM is best suited when the aim is an interpretable, descriptive fuzzy partition of the array, whereas generative and mixture models are preferable when statistical inference on latent memberships is required, and spectral or factorization-based methods when scalability is the main concern.

\section{Simulation study}\label{sec4}

A simulation study has been implemented to assess the performance of the proposed method for each of eight experimental conditions, obtained by crossing
two data sizes, two levels of cluster complexity and two noise levels. 

Objects, variables and occasions are assigned at random to $P$, $Q$, $R$ clusters through binary membership matrices $\mathbf{U}$, $\mathbf{V}$ and
$\mathbf{W}$, with the constraint that no cluster is empty. The core array $\underline{\mathbf{G}}$ is generated by drawing its entries from a standard
normal distribution, $g_{pqr}\sim\mathcal{N}(0,1)$, and the noiseless signal array is $\underline{\mathbf{X}}^{s}=[g_{z_i y_j t_k}]$, where $z_i$, $y_j$,
$t_k$ are the true cluster labels. A noise array $\underline{\mathbf{E}}$ with
independent $\mathcal{N}(0,1)$ entries is then generated.
 
Both $\underline{\mathbf{X}}^{s}$ and $\underline{\mathbf{E}}$ are normalized
across the three modes, and the same normalization is applied to the final array, so that scale differences among objects, variables and occasions do not
drive the partitions. The normalization is computed iteratively: at each iteration the array is
first centered by subtracting the object, variable and occasion means and adding back twice the
grand mean, so that all slab means are zero, and then each slab of each mode is divided in turn by
its root mean square. These two steps are repeated until the maximum deviation of the slab means
from 0 and of the slab variances from 1 falls below $10^{-10}$ (with a maximum of 5000 iterations). The observed array is therefore $\underline{\mathbf{X}}=\mathcal{N}\!\left(\mathcal{N}(\underline{\mathbf{X}}^{s})+\sigma\,\mathcal{N}(\underline{\mathbf{E}})\right)$. Because signal and noise both have unit variance in every slab before mixing, $\sigma$ is the noise-to-signal standard deviation ratio.

Two data sizes are considered: a small size with $I=40$, $J=30$, $K=20$ and a large size with $I=100$, $J=75$, $K=50$. Each size is combined with two numbers of clusters: a low-complexity configuration with $P=4$, $Q=3$, $R=2$ and a high-complexity configuration with $P=5$, $Q=4$, $R=3$. This gives four design cells, denoted Small--Small, Small--High, Large--Small and Large--High, where the first term refers to the data size and the second to the number of clusters. Each cell is run at a low ($\sigma=3$) and a high ($\sigma=7$) noise
level, for a total of eight scenarios. Since clusters for each dimension have been generated with the same variance around centroids, we consider $\lambda_{\mathbf{U}}=\lambda_{\mathbf{V}}=\lambda_{\mathbf{W}}$ for
coherence. 

\begin{figure}[ht!]
\centering
\includegraphics[scale=0.35]{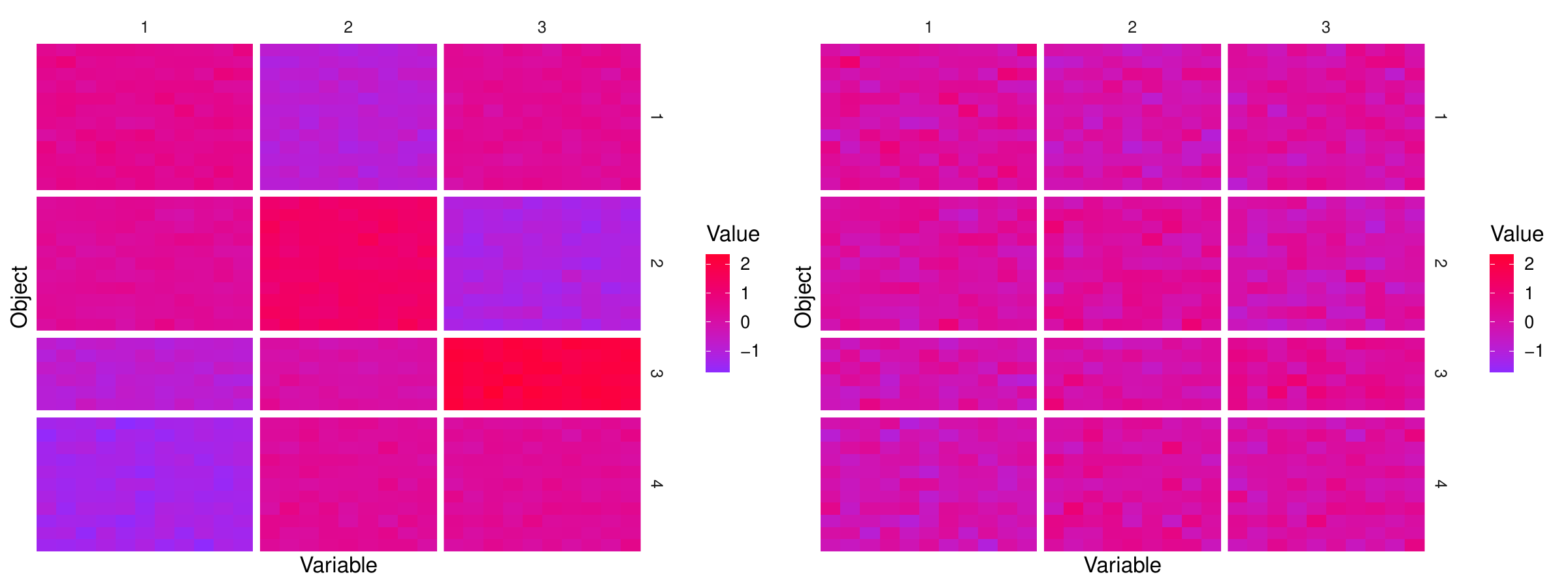}

\caption{Examples of simulated samples for the Small-Small scenario, for low (left) and high (right) error. Object-Variable clusters heatmaps are obtained averaging over the occasions in Cluster 1}
\label{fig1}       
\end{figure}

From the two examples of generated data samples in Figures \ref{fig1} (and Figure and Figure \ref{fig1.0} in Appendix~\ref{app:example}), it is possible to observe that the simulated groups are clearly visible when a small error is considered, and tend to be less visible as the error grows.

Triple fuzzy entropic $k$-means algorithm (accelerated version) has been applied to recover the true partition of each way, quantifying also the associated degree of fuzziness. In all scenarios the numbers of clusters $(P,Q,R)$ were treated as known and set equal to
their true values. Its performance has been evaluated by calculating the average over \textit{nsim} $= 100$ of the following measures for each scenario:

\vspace{0.1cm}

\begin{enumerate}[itemsep = -0.5pt]
    \item for each way, the mean, the median and the quartiles $(Q_1,Q_3)$ of the Adjusted Rand Index (ARI, \cite{hubert1985comparing}) between the generated crisp membership matrices and the ones obtained by taking the maximum membership of each row of the estimated fuzzy matrices $\hat{\mathbf{U}}$, $\hat{\mathbf{V}}$, and $\hat{\mathbf{W}}$;
    \item the discrepancy between the generated centroids $\mathbf{G}^*$ and the estimated ones $\hat{\mathbf{G}}$ is evaluated considering the percentage sum of squared residuals over the total sum of squares 

    \begin{equation}
    \frac{||\hat{\underline{\mathbf{G}}} - \underline{\mathbf{G}^*}||^2}{||\underline{\mathbf{G}^*}||^2} = \frac{\sum_{p=1}^{P}\sum_{q=1}^{Q}\sum_{r=1}^{R} (\hat{g}_{pqr} - g_{pqr}^*)^2}{\sum_{p=1}^{P}\sum_{q=1}^{Q}\sum_{r=1}^{R} (g_{pqr}^*)^2};
    \label{eq:discrepancy}
    \end{equation}    
    \item average number of iterations needed for convergence;
    \item average number of seconds require to reach convergence from an initial starting point;
    \item chosen value of the fuzziness parameter $\lambda$.
\end{enumerate}

It is worth stressing that, to correctly compare the generated core array $\underline{\mathbf{G}}^*$ with the estimated one $\hat{\underline{\mathbf{G}}}$ using the discrepancy measure defined in (\ref{eq:discrepancy}), it is necessary to solve a linear sum assignment problem. This accounts for the label switching issue inherent to many clustering algorithms, where the estimated clusters may be permuted relative to the true ones. Consider, for instance, the case with $P = 2$ clusters for the object way. It may happen that the algorithm perfectly estimates the generated centroid for cluster 1, but assigns it to the estimated cluster 2. To manage this possible switching, we first construct a vector $\mathbf{t}$ of length $P$ containing the true cluster labels (ordered according to the row norms of the matricized true centroid array) and a vector $\mathbf{f}$ of length $P$ containing the estimated cluster labels (similarly ordered based on the estimated centroids). A confusion matrix $\mathbf{S} = \mathbf{U}'_{true}\hat{\mathbf{U}}$ of size $P \times P$ is then built, where $\mathbf{U}_{true}$ is the true partition and $\hat{\mathbf{U}}$ the partition estimated by the algorithm. Then the permutation maximizing agreement, is found,  This assignment is found in $O(P^3)$ time using the primal-dual Hungarian method \citep{bib4}. A detailed discussion of algorithms for solving the assignment problem can be found in \citep{bib19}. Once the optimal permutation is obtained, an analogous procedure is performed for variables and occasions. Then, all three permutations consistently to the estimated centroid array are applied $\hat{\underline{\mathbf{G}}}$ before computing the discrepancy measure.
\begin{table}[ht!]
\caption{Simulation results: mean values, standard deviations in parentheses}
\label{tab2}
\centering
\setlength{\tabcolsep}{5pt}
\renewcommand{\arraystretch}{1.1}

\begin{tabular}{lcc|cc}
\cline{1-5}
& \multicolumn{4}{c}{\textbf{Small data size}} \\
\cline{2-5}
 & \multicolumn{2}{c|}{\textbf{Small \# clusters}}
& \multicolumn{2}{c}{\textbf{Large \# clusters}} \\
\cline{2-5}
& Low error & High error & Low error & High error \\
\hline
\textbf{Mean ARI}$_{\mathbf{U}}$
& 0.992 (0.037) & 0.673 (0.314) & 0.968 (0.125) & 0.275 (0.231) \\
\textbf{Mean ARI}$_{\mathbf{V}}$
& 0.999 (0.010) & 0.739 (0.378) & 0.973 (0.117) & 0.346 (0.280) \\
\textbf{Mean ARI}$_{\mathbf{W}}$
& 1.000 (0.000) & 0.803 (0.370) & 0.977 (0.144) & 0.398 (0.351) \\
\textbf{Centroids disc.}
& 0.008 (0.080) & 0.311 (0.431) & 0.097 (0.254) & 0.958 (0.349) \\
\textbf{Iter.}
& 5.610 (1.270) & 9.760 (4.971) & 6.410 (1.865) & 8.630 (5.220) \\
\textbf{Time}
& 3.057 (0.707) & 5.251 (1.867) & 3.327 (0.708) & 4.214 (1.642) \\
\textbf{Chosen} $\lambda$
& 0.297 (0.208) & 0.285 (0.173) & 0.266 (0.182) & 0.224 (0.176) \\
\cline{1-5}
& \multicolumn{4}{c}{\textbf{Large data size}} \\
\cline{2-5}
\textbf{Mean ARI}$_{\mathbf{U}}$
& 1.000 (0.000) & 0.881 (0.257) & 0.998 (0.024) & 0.820 (0.281) \\
\textbf{Mean ARI}$_{\mathbf{V}}$
& 1.000 (0.000) & 0.897 (0.248) & 1.000 (0.000) & 0.819 (0.279) \\
\textbf{Mean ARI}$_{\mathbf{W}}$
& 1.000 (0.000) & 0.965 (0.159) & 1.000 (0.000) & 0.804 (0.363) \\
\textbf{Centroids disc.}
& 0.000 (0.000) & 0.194 (0.394) & 0.006 (0.056) & 0.355 (0.466) \\
\textbf{Iter.}
& 4.680 (0.777) & 5.760 (1.865) & 4.590 (0.866) & 6.520 (2.851) \\
\textbf{Time}
& 15.646 (2.987) & 20.585 (3.934) & 14.268 (2.667) & 23.240 (4.238) \\
\textbf{Chosen} $\lambda$
& 0.355 (0.204) & 0.327 (0.208) & 0.264 (0.163) & 0.122 (0.194) \\
\hline
\end{tabular}
\end{table}
Analyzing the results in Table~\ref{tab2} and in the green boxplots of Figure~\ref{boxplot}, for the small-size scenarios with low error the clustering accuracy, measured through the ARI, is excellent in all three ways
(mean ARI between 0.968 and 1.000), and the estimated centroids are very close to the true ones. As expected, performance declines under high noise, and the decline depends strongly on cluster complexity. In the Small--Small design the mean ARI drops to 0.67--0.80 and the centroid discrepancy to 0.311, so the core structure is still partly recovered. In the Small--High design,
instead, the mean ARI falls to 0.28--0.40 and the centroid discrepancy reaches 0.958, i.e., with many clusters, a small array and high noise, neither the
partitions nor the centroids are reliably recovered. The algorithm converges in about 4.6 to 9.8 iterations on average, with more iterations under high noise,
and average runtimes of about 3 to 5 seconds in the small scenarios. 

In the large-size scenarios, the results for low error are very close to those of the small-size scenarios. Thus, larger arrays make the recovery less sensitive to
noise, while a larger number of clusters makes the recovery harder in all settings. The computational cost increases with the array size, with average
runtimes of about 14 to 23 seconds in the large scenarios, whereas the number of iterations does not increase (4.6 to 6.5 on average).

\subsection{Lambda mis-specification}\label{mis}

To further validate the importance and effectiveness of the proposed procedure for selecting the optimal fuzziness parameter $\lambda$, an additional robustness experiment was conducted. The aim was to demonstrate that selecting $\lambda$ inappropriately, i.e. by ignoring the underlying variability structure of the data, leads to a deterioration in clustering performance.

\begin{figure}[ht!]
\centering
\includegraphics[scale=0.35]{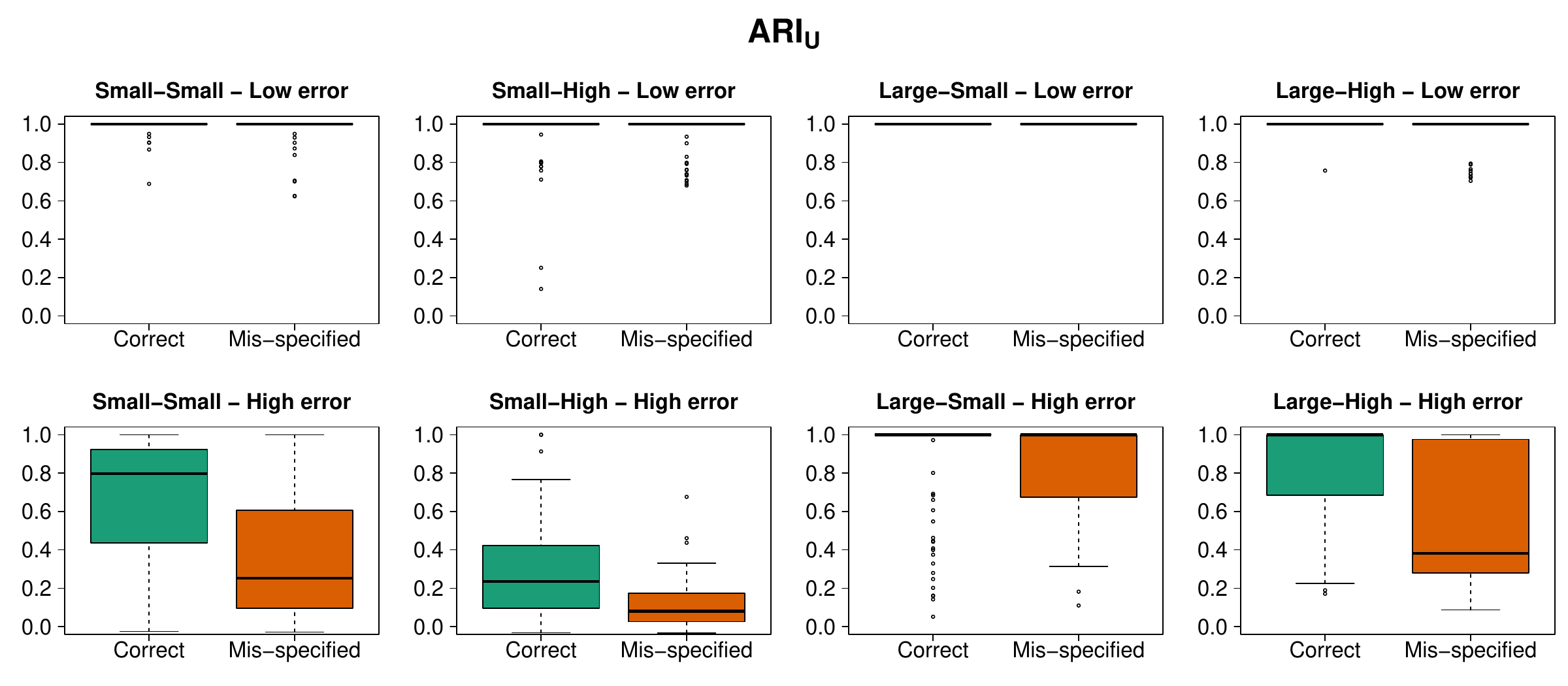}
\caption{Boxplots of ARI$_\mathbf{U}$ values under calibrated and mis-specified $\lambda$}
\label{boxplot}
\end{figure}

In this experiment, we intentionally not optimize the  $\lambda$ value (Mis-specified), considering a default of 0.5. The rationale is that if our model's $\lambda$ selection procedure is truly adaptive and data-sensitive, then deviating from its recommendations should negatively impact clustering accuracy. The clustering algorithm was re-run with these mis-specified $\lambda$ values, keeping all other settings unchanged. Figure \ref{boxplot} presents boxplots of the units ARI for each data way under the optimally specified and mis-specified $\lambda$ configurations. Results for variables and occasions can be found in Appendix~\ref{app:mis}.

With low error, the calibrated and mis-specified
configurations are nearly indistinguishable: the median ARI equals 1 in all scenarios and ways. Thus, when the signal-to-noise ratio is high, the recovery is robust to the value of $\lambda$. With high error, using $\lambda = 0.5$ instead of the calibrated value has a clear cost. For instance, in Small-Small scenario, the median object ARI falls from about 0.80 to about 0.25, and in Large-High scenario from 1 to about 0.4, with a much wider interquartile range. The decline is similar for variables and occasions. 

\subsection{Comparison with tandem and independent procedures}\label{subsec:comparison}
 
\begin{figure}[ht!]
\centering
\includegraphics[width=\textwidth]{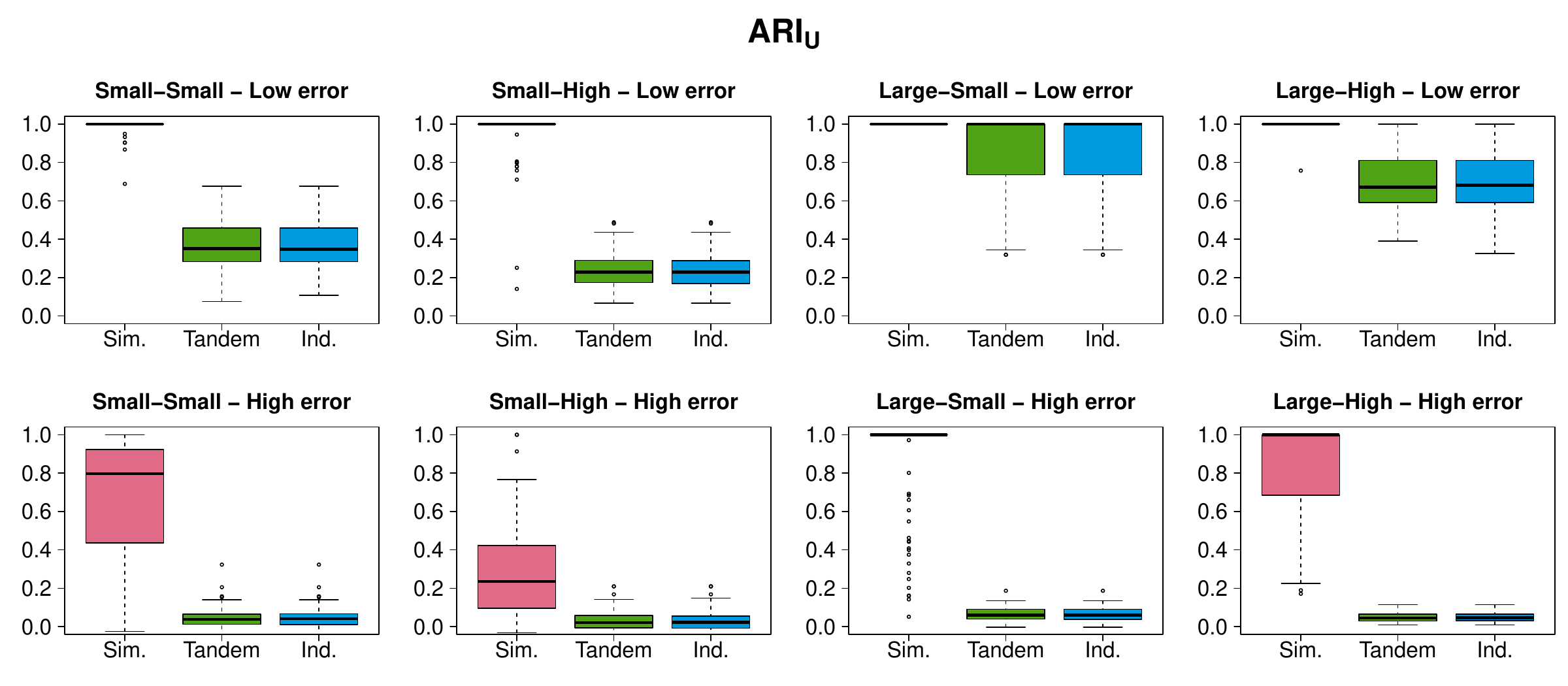}
\caption{Boxplots of ARI$_{\mathbf{U}}$ for the simultaneous (Sim.), tandem (Tandem) and independent (Ind.) procedures. Panels are labeled as data size--number of clusters -- noise level}
\label{fig:cmp_U}
\end{figure}

To assess the benefit of estimating the three partitions simultaneously, FE3KM (Sim.) is compared, on the same simulated data sets and in all eight scenarios, with two alternatives. The first is the tandem procedure of
Subsection~\ref{tandem} (Tandem), in which fuzzy entropic $k$-means is applied sequentially to objects, variables and occasions, each step being conditioned on the centroids estimated in the previous one. The second is independent clustering (Ind.) of Subsection \ref{sec:independent}, in which fuzzy entropic $k$-means is applied separately to the three matricized arrays $\mathbf{X}_{I,JK}$, $\mathbf{X}_{J,IK}$ and
$\mathbf{X}_{K,IJ}$, so that each partition ignores the other two modes. The two comparisons therefore isolate the effect of conditioning on previously
estimated partitions (Sim.\ vs.\ Tandem) and the effect of disregarding the cross-mode structure altogether (Sim.\ vs.\ Ind.). Performance is measured by
the ARI between the true and the estimated partition of each way, computed on $\textit{nsim} = 100$ replications per scenario. Figure~\ref{fig:cmp_U} reports the
results for the objects (ARI$_{\mathbf{U}}$); those for variables (ARI$_{\mathbf{V}}$) and occasions (ARI$_{\mathbf{W}}$) are reported in
Appendix~\ref{app:comparison}.
 
For the object way, the simultaneous procedure outperforms both competitors. With low error, FE3KM recovers the true partition almost perfectly in all four design cells, with median ARI equal to 1 and only a few low outliers. Tandem and Ind.\ perform substantially worse in the Small scenarios, with median ARI values around 0.35 in Small--Small and 0.2 in Small--High, whereas their performance improves in the Large cells. With high error, the advantage of FE3KM becomes more pronounced: Tandem and Ind.\ have median ARI values close to zero in every scenario, while FE3KM retains a median ARI of approximately 0.8 in Small--Small and 1 in both Large cells. Small--High remains the most challenging condition, with the FE3KM median also declining substantially, although it remains above those of both competitors. These results confirm that larger arrays improve object-partition recovery under noise, whereas increasing the number of clusters makes recovery more difficult. The results for \textbf{V} and \textbf{W} (Appendix~\ref{app:comparison}) lead to similar conclusions.

\section{Application}\label{sec5}

The potential of the proposed methodology is further shown through the application on the well-known three-way TV dataset \citep{lundy89}. It refers to the ratings on $J =$ 15 American TV programs with respect to $I = $ 16 bipolar scales given by a group of $K =$ 30 psychology students at the University of Western Ontario. The three ways correspond to bipolar rating scales (objects), TV programs (variables), and  students (occasions). The FE3KM model provides three simultaneous fuzzy partitions, which we interpret as follows: ($i$) a clustering of rating scales into higher-order dimensions of evaluation, ($ii$) a grouping of TV shows with similar perceptual patterns across scales and students, and ($iii$) a segmentation of students according to their overall viewing profiles. 

The numbers of clusters $(P,Q,R)$ are selected by means of a three-way extension of the two-way Xie--Beni (XB) index proposed by \cite{ferraro2026fuzzy}. As in the two-way case, the index is the ratio between a measure of compactness and a measure of separation. Compactness is the average within-block deviance of the fitted solution,
\begin{equation}
  c = \frac{1}{IJK}\sum_{i=1}^{I}\sum_{j=1}^{J}\sum_{k=1}^{K}\sum_{p=1}^{P}\sum_{q=1}^{Q}\sum_{r=1}^{R} u_{ip}\,v_{jq}\,w_{kr}\,(x_{ijk}-g_{pqr})^2 = \frac{WSS}{IJK},
\end{equation}
where the entropy-regularized memberships play the role of the fuzzified memberships, so that no explicit fuzzifier is needed. Separation is measured, for each mode, by the minimum weighted squared distance between two blocks of centroids differing in that mode only:
\begin{align}
  s_\mathbf U &= \min_{p\neq p'} \sum_{qr} \omega_{qr}\,(g_{pqr}-g_{p'qr})^2, \\
  s_\mathbf V &= \min_{q\neq q'} \sum_{pr} \omega_{pr}\,(g_{pqr}-g_{pq'r})^2, \\
  s_\mathbf W &= \min_{r\neq r'} \sum_{pq} \omega_{pq}\,(g_{pqr}-g_{pqr'})^2,
\end{align}
with weights $\omega_{qr}=\frac{v_{+q}}{J}\frac{w_{+r}}{K}$, $\omega_{pr}=\frac{u_{+p}}{I}\frac{w_{+r}}{K}$ and $\omega_{pq}=\frac{u_{+p}}{I}\frac{v_{+q}}{J}$, each set summing to one. The index is then
\begin{equation}
  XB(P,Q,R) = \frac{c}{\min\left(s_\mathbf U,\,s_\mathbf V,\,s_\mathbf W\right)},
\end{equation}
and the optimal configuration is the one that minimizes $XB$ over the candidate triplets $(P,Q,R)$, since low values indicate compact and well-separated blocks. For each candidate triplet, the regularization coefficients are tuned by cross-validation and the model is refitted on the full array before the index is computed.


Table~\ref{tab:XB} in Appendix~\ref{app:xb} reports the attained $XB$ values. The global minimum ($XB$ $=7.7$) is attained at $(P,Q,R)=(3,6,2)$, so the selected solution has 3 clusters of rating scales, 6 clusters of TV programs and 2 clusters of students. Values below $10^{2}$ occur only with two student clusters and two or three scale clusters, while all configurations with $R\ge 3$ attain values larger than $10^{4}$.

The optimal lambdas obtained through cross-validation where \\
$\lambda_{scale} = 0.38,\ \lambda_{program} = 0.30,\ \lambda_{student} = 0.13$. 

\begin{figure}[ht!]
\centering
\includegraphics[scale=0.55]{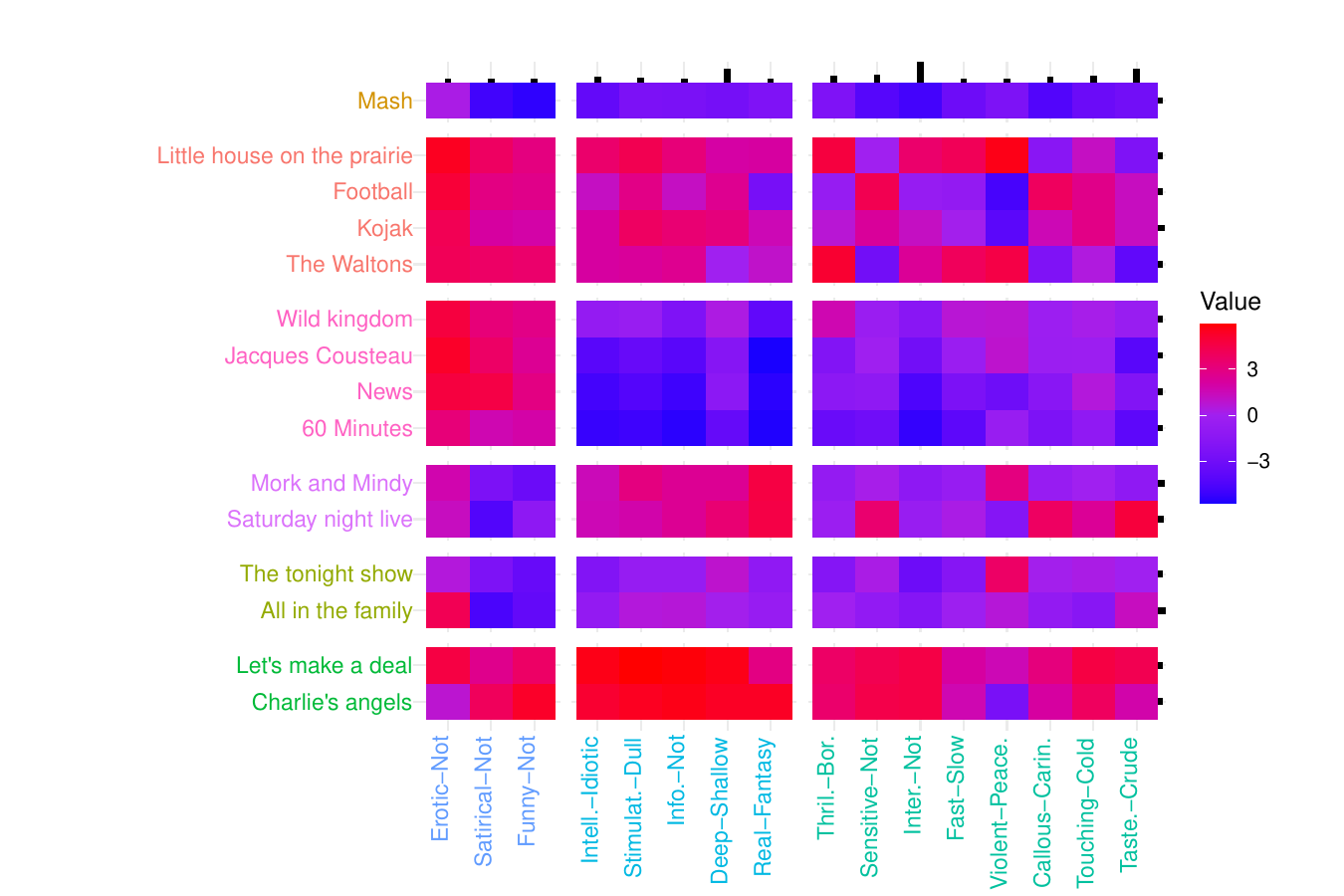}
\caption{Clustering of the scales and of the TV shows, each point is the average over the students in Cluster 1}
\label{c1}       
\end{figure}

\begin{figure}[ht!]
\centering
\includegraphics[scale=0.55]{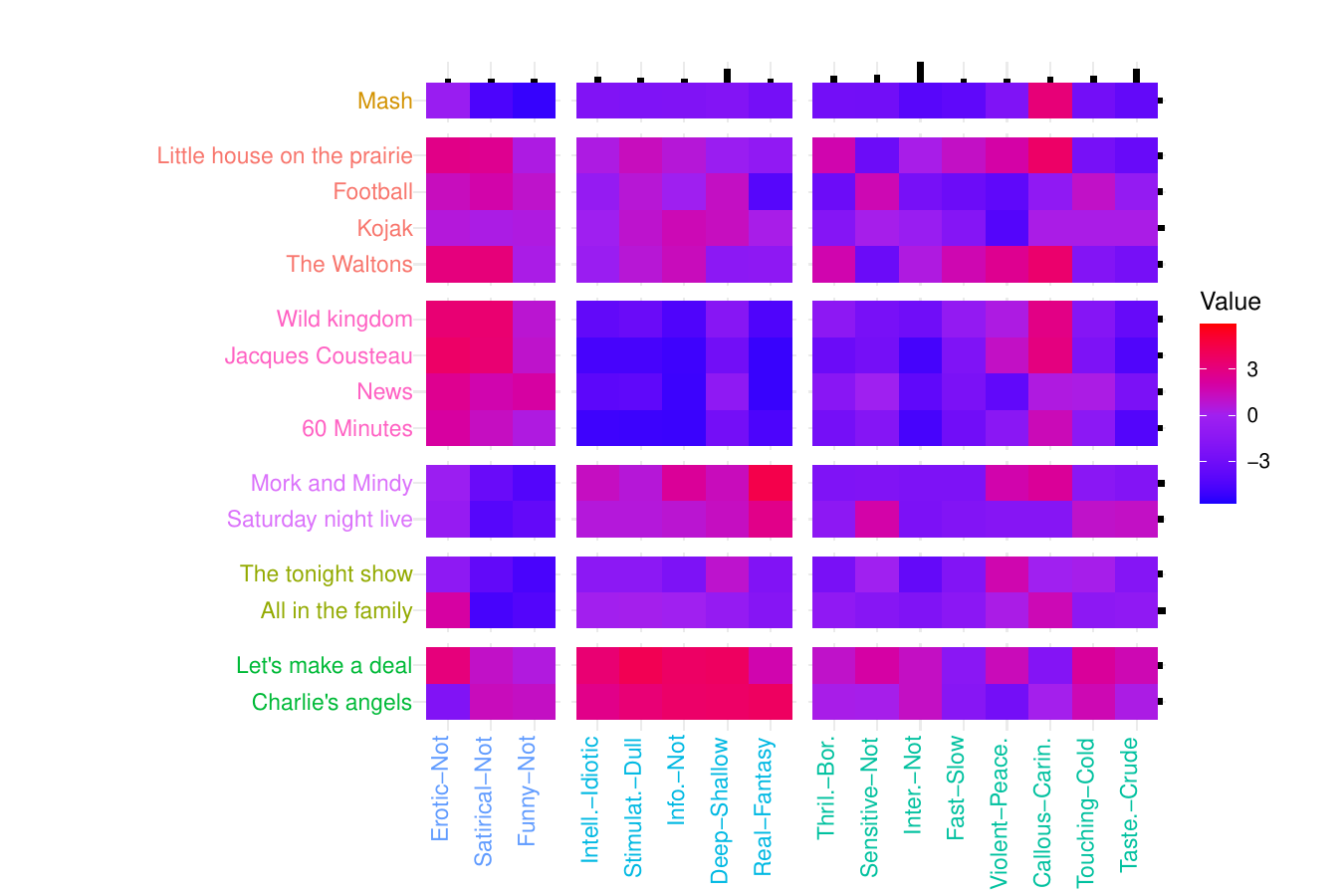}
\caption{Clustering of the scales and of the TV shows, each point is the average over the students in the Cluster 2}
\label{c2}       
\end{figure}

Figures \ref{c1} and \ref{c2} show the clusters of TV shows and scales averaged for the each of the two clusters of students that we have found. The horizontal and vertical black bars indicate the uncertainty associated with the assignment of that entity to its cluster. This uncertainty is computed, for instance for the units but analogously for the other ways, as $(1-\max_p u_{ip})/(1-1/P)$, ranging from 0 to 1, where 0 corresponds to a perfectly crisp assignment and 1 to maximum uncertainty. Hence, entities with small bars are strongly and consistently associated with their cluster, while those with larger bars exhibit weaker memberships across clusters. These bars highlight the added value of fuzzy clustering compared with hard clustering: instead of forcing every entity into a single group, the FE3KM solution quantifies how confidently each belongs, revealing degrees of membership that hard partitions conceal. This is particularly informative for entities that are heterogeneous or share features with multiple clusters. 

The displayed heatmaps are coherent with this interpretation: entities whose patterns do not align cleanly with a single cluster show higher uncertainty values, whereas those forming clear, homogeneous blocks exhibit low uncertainty. 

Figures~\ref{c1} and~\ref{c2} show that the two student clusters differ in how judgments are expressed. Students in Cluster~1 use the rating scales more extensively, with more extreme values and sharper contrasts between programs, and they lean towards the second pole of most bipolar scales. In particular, they rate Little House on the Prairie, Football, Kojak, The Waltons and the nature and news programs as clearly not erotic, not satirical and not funny; they find Let's make a deal and Charlie's angels uninteresting, insensitive and crude; and they judge Little house on the prairie and The Waltons boring, slow and peaceful. Students in Cluster~2 give more moderate ratings: their values stay closer to the midpoint on the humor and erotic scales for these programs and, on the remaining scales, tend more towards the first pole (thrilling, sensitive, interesting). They also give a more emotional reading of the family and nature programs (Little house on the prairie, The Waltons, Wild kingdom, Jacques Cousteau), which they find more touching and caring.

For what concerns the clustering of the scales, the three clusters can be interpreted as follows. Cluster 1 groups the scales concerning humor, satire and erotic content, i.e., the dimension that separates comedic and provocative programs from the others. Cluster 2 gathers the scales related to intellectual and cognitive engagement and to realism. Cluster 3 is the largest and it can be read as an affective--evaluative dimension. These results can be related to previous works that applied dimensionality reduction to this dataset \citep{giordani14}, which identified three principal components for the scales: \lq\lq Humor\rq\rq, \lq\lq Sensitivity\rq\rq\ and \lq\lq Violence\rq\rq. The correspondence is not one-to-one, since clusters are not components. Cluster 1 recovers the comedic and satirical side of \lq\lq Humor\rq\rq, together with eroticism, whereas its cognitive aspects (e.g., intellectual stimulation) are separated and form a cluster of their own (Cluster 2). The \lq\lq Sensitivity\rq\rq\ and \lq\lq Violence\rq\rq\ components, instead, are not separated and are merged into Cluster 3. The scales with the largest uncertainty, \lq\lq Interesting--Uninteresting\rq\rq\ and \lq\lq Tasteful--Crude\rq\rq, belong to this cluster, which suggests that they partly overlap with the other two. 

We now focus on the clustering of the TV shows, whose six clusters are numbered from top to bottom as displayed in Figures~\ref{c1} and~\ref{c2}. Cluster 2 groups Little House on the Prairie, Football, Kojak and The Waltons: these programs take extreme and opposite values on the excitement and violence scales in Cluster 3. Indeed, Football and Kojak are violent and callous, and to a lesser extent thrilling and fast, while The Waltons and Little house on the prairie are peaceful, slow and boring. Cluster 3 (Wild Kingdom, Jacques Cousteau, News, 60 Minutes) shows the lowest values on the intellectual scales in Cluster 2 and high values on the scales in Cluster 1: these are the most real, informative and intelligent programs, and the least funny, satirical or erotic. Cluster 4 (Mork and Mindy, Saturday night live) shows the lowest values on the humor and satire scales in Cluster 1, so it contains the funniest and most satirical shows, together with slightly high values on the intellectual scales in Cluster 2, with Mork and Mindy being the program rated most as fantasy. Cluster 5 (The Tonight Show, All in the Family) is also comedic, but with a more substantive profile: these programs are judged less idiotic and less shallow than those in Cluster 4, and more interesting. Finally, Cluster 6 (Let's make a deal, Charlie's angels) takes the highest values on the intellectual scales in Cluster 2 (idiotic, dull, uninformative, shallow, fantasy) and high values on the scales in Cluster 3 (uninteresting, insensitive, cold), i.e., these are the least engaging programs.

Mash, which forms Cluster 1 by itself, is the program that most strongly combines very low values on the humor and satire scales (it is judged very satirical and funny) with low values on the intellectual scales (it is judged intelligent, informative and real) and on the emotional scales in Cluster 3, where it is also judged caring. This is consistent with the dual nature of Mash: it is both a realistic depiction of wartime conditions and a sharp social satire infused with humor, which is why it cannot be grouped with either the documentary programs of Cluster 3 or the comedy programs of Clusters 4 and 5. 

Finally, the two student clusters differ mainly in the intensity of their ratings rather than in the structure of the blocks. Considering, for instance, the scale \lq\lq Thrilling--Boring\rq\rq\ for the TV shows in Cluster 2, students in Cluster 1 find The Waltons and Little House on the Prairie much more boring than students in Cluster 2, whereas students in Cluster 2 find Football and Kojak more thrilling than students in Cluster 1.

Figure~\ref{fig:BSS_comp} displays cluster-level contributions ($B_p^{(\mathbf U)}/TSS$, $B_q^{(\mathbf V)}/TSS$, $B_r^{(\mathbf W)}/TSS$) on the total-variation scale (Proposition 5):
\begin{figure}[ht!]
    \centering
    \includegraphics[width=0.65\linewidth]{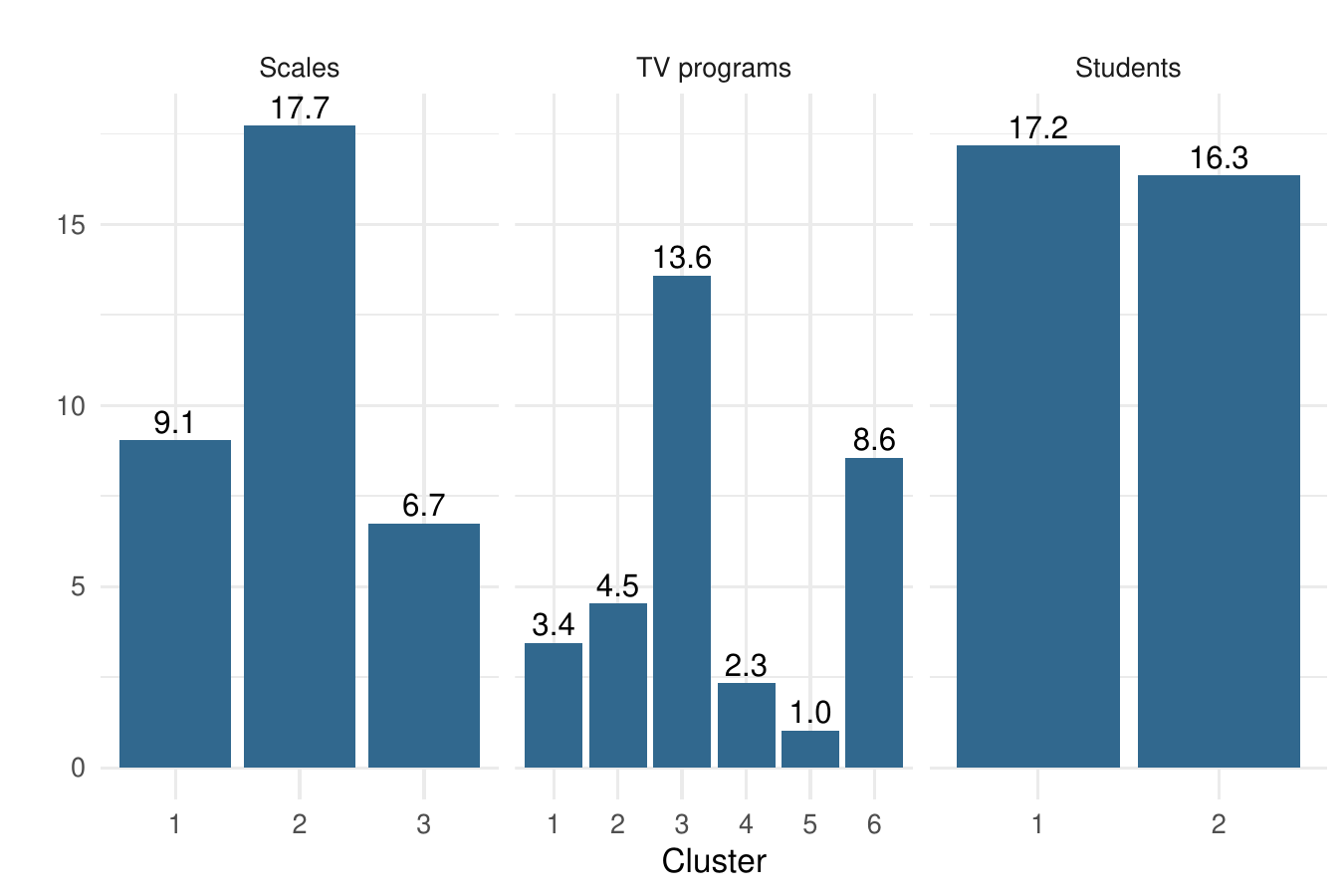}
    \caption{Contribution of each cluster across the three ways (Scales, TV programs, and Students) to the Total Sum of Squares ($\% \text{ of } B_.^{(.)}/ TSS$). Normalized contributions within each way sum to $BSS = 33.5\%$}
    \label{fig:BSS_comp}
\end{figure}
For what concerns the scales, Cluster~2 (intellectual and cognitive engagement scales) provides the largest individual contribution ($17.7\%$ of $TSS$), followed by Cluster~1 (humor, satire and erotic content, $9.1\%$), while Cluster~3 (affective--evaluative scales) accounts for the smallest percentage ($6.7\%$), even though it is the largest cluster: its blocks lie close to the overall mean, so these scales contribute less to the separation among blocks. Programs in Cluster~3 (highly informative and intelligent programs: Wild kingdom, Jacques Cousteau, News, 60 Minutes) dominate the program way with $13.6\%$ of $TSS$, followed by Cluster~6 (Let's make a deal, Charlie's angels, $8.6\%$), whereas Cluster~2 and Cluster~1 (Mash) contribute $4.5\%$ and $3.4\%$, and the comedy-oriented Clusters~4 and~5 only $2.3\%$ and $1.0\%$. This highlights that the contrast between informative, documentary content and the least engaging programs, which lie at the two extremes of the intellectual scales, generates the strongest separation among programs, consistently with the largest contribution of Cluster~2 among the scales. Finally, student Cluster~1 accounts for $17.2\%$ of $TSS$, while Cluster~2 accounts for $16.3\%$ of $TSS$, demonstrating a balanced distribution of explained variance across the two viewer segments.

Moreover, to quantify the isolated explanatory gain from partitioning a single way while holding the remaining two membership matrices fixed, we compute the conditional deviance gains $\Delta$ illustrated in Proposition 6. Expressed as a proportion of total variation ($\Delta / TSS$), these conditional gains measure the drop in explained sum-of-squares when a specific mode is pooled into a single cluster:
\begin{itemize}[label=-]
    \item \textbf{TV Programs ($\Delta^{(\mathbf V)} / TSS = 29.02\%$)}: partitioning TV programs generates the highest conditional gain, demonstrating that program differentiation is the primary driver of structure in the evaluation dataset.
    \item \textbf{Scales ($\Delta^{(\mathbf U)} / TSS = 14.17\%$)}: the scale partition yields a conditional gain of $14.17\%$, confirming distinct baseline responses across perceptual dimensions.
    \item \textbf{Students ($\Delta^{(\mathbf W)} / TSS = 3.40\%$)}: the student partition yields a modest conditional gain, indicating that individual viewer variation provides a smaller incremental gain compared to program type and scale distinctions.
\end{itemize}

Proposition~2 shows that a multi-stage sequential tandem clustering procedure decomposes the joint loss into additive stage-specific terms $f_{FE3KM}(\mathbf{U}, \mathbf{V}, \mathbf{W}, \mathbf{G}) = f_1(\mathbf{U}, \mathbf{M}) + f_2(\mathbf{V}, \mathbf{C}) + f_3(\mathbf{W}, \mathbf{G})$. 
\begin{figure}[ht!]
    \centering
    \includegraphics[width=0.95\linewidth]{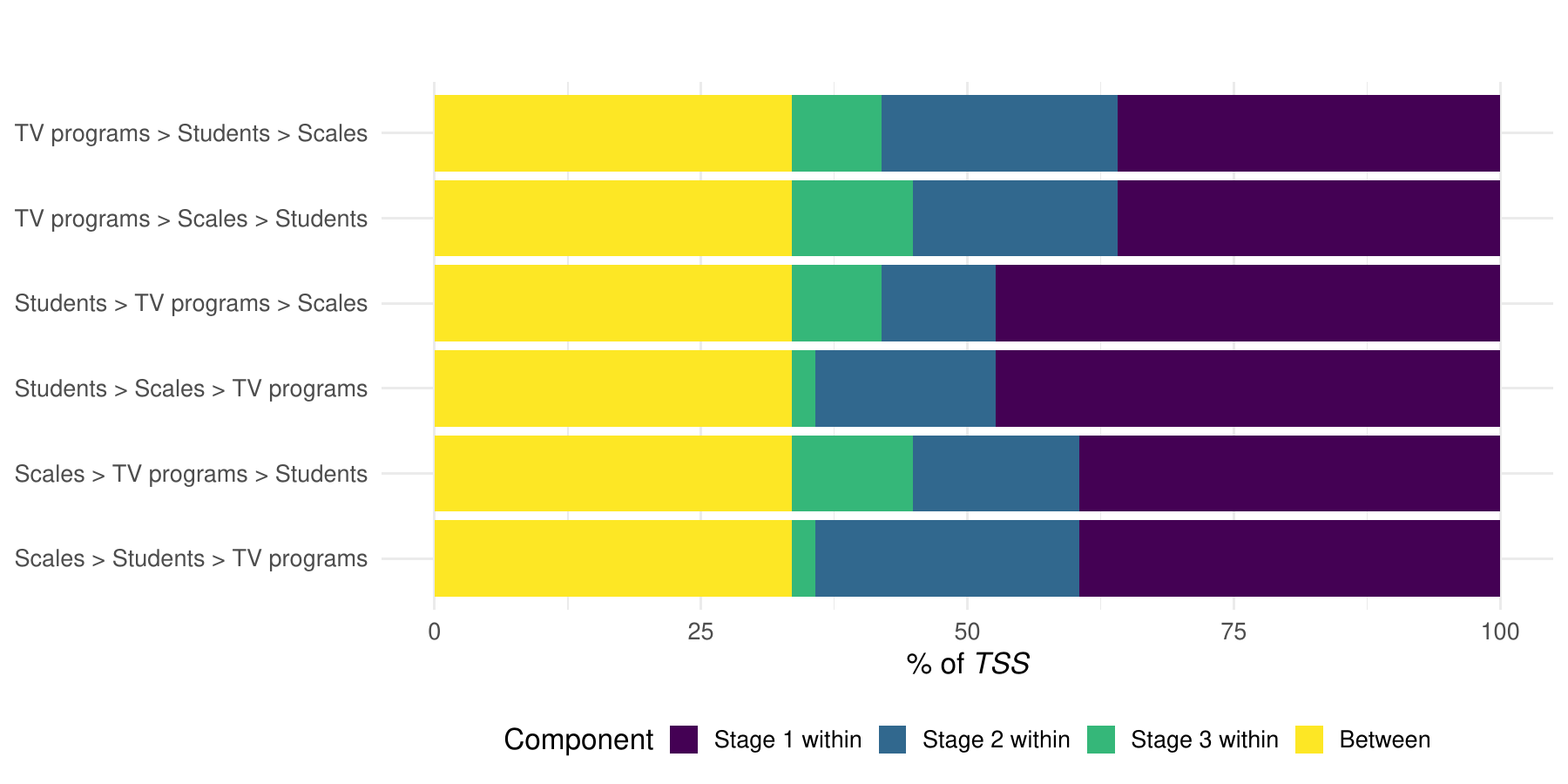}
    \caption{Tandem decomposition of the final optimal solution. While overall $WSS$ ($66.5\%$) and $BSS$ ($33.5\%$) remain invariant across permutations, the relative size of each intermediate stage's within-cluster loss depends on the ordering sequence}
    \label{fig:tandem_decomp}
\end{figure}
Figure \ref{fig:tandem_decomp} decomposes the unregularized within-block sum of squares of the same FE3KM solution under six mode orders. The intermediate allocations depend on the order, whereas total $WSS$ and $BSS$ remain unchanged. It illustrates why tandem methods are vulnerable to step-wise greedy path bias, motivating the simultaneous optimization approach of FE3KM. 

To isolate how explained variance is distributed across main effects and multi-way interaction terms, an ANOVA-type decomposition of $BSS$ is also performed. The breakdown reveals:
\begin{itemize}[label =-]
    \item Main effect of TV Programs ($47.6\%$ of $BSS$): program distinctions constitute the single largest main effect, accounting for nearly half of the explained variance.
    \item Scales $\times$ TV Programs Interaction ($36.8\%$ of $BSS$): the strong two-way interaction demonstrates that viewers evaluate different program genres along distinct perceptual scale dimensions.
    \item Main effect of Students ($8.0\%$ of $BSS$): captures moderate baseline evaluation shifts between student groups. 
    \item Main effect of Scales ($5.4\%$ of $BSS$): reflects overall baseline differences across rating scale items.
    \item Other interactions ($2.1\%$ total): Student interactions, i.e. TV programs $\times$ Students ($2.0\%$), Scales $\times$ Students ($0.0\%$), and three-way interaction ($0.1\%$) are minimal.
\end{itemize}

\section{Conclusions and final remarks}\label{sec6}

We introduced a novel methodology for fuzzy multi-way partitioning of three-way data arrays, expanding the flexibility and interpretability of fuzzy clustering in complex data structures. Our framework generalizes classical models: by setting penalty components to zero, the model reduces to a hard clustering of objects, variables, and occasions. Further degenerating the model to, for instance, object-only partitioning, the loss recovers the well-known $k$-means algorithm, placing our approach within a broader methodological context.
It has been shown how to estimate the three-way partitioning model by using tandem procedures based on the fuzzy entropic $k$-means algorithm. These procedures generally give sub-optimal estimations of the three-way model. Of course, other tandem procedures can be easily derived from the simultaneous approach. In particular, partitioning of one way + co-clustering of the other two (or vice versa) represent intermediate strategies between the fully tandem and the fully simultaneous cases. Building on this a new model, named fuzzy entropic triple $k$-means, has been developed and analyzed, together with Alternating Least-Squares algorithms for its estimation. The model simultaneously represents the three-way partitioning model under minimal assumptions using a Least-Squares approach. Moreover, it has been shown that the objective function of the simultaneous approach can be rewritten as the sum of the three objective functions of the tandem approach.

Comprehensive experiments on synthetic and real datasets validated the method, demonstrating accurate recovery of underlying three-way clustering structures. Importantly, the penalty terms, tuned via cross-validation, adaptively control cluster fuzziness. Note that, when data are homogeneous, there is no need to choose different penalties for different partitions, which simplifies computations.

The application to the benchmark TV shows dataset showcased not only the effectiveness of our approach, but also its potential for insightful interpretations of real-world scenarios. By combining methodological rigor with practical applicability, this work significantly advances multi-way clustering techniques for analyzing complex three-way data. Future research directions include extending these methods to higher-order data arrays and exploring alternative loss functions to further enhance flexibility and robustness. A further
direction concerns the choice of the numbers of clusters $(P,Q,R)$. In this paper they were treated as known in the simulation study and selected with a three-way Xie--Beni index in the application. A systematic study of the sensitivity of this choice to the data, the noise level and the fuzziness parameters would be an interesting development for three-way fuzzy clustering. Overall, the proposed framework equips researchers and practitioners with powerful tools to uncover complex latent structures in multidimensional data.

\section*{Declarations}

\textbf{Funding and conflict of interest}. The authors did not receive support from any organization for the submitted work, and the authors have no financial or non-financial conflict of interest to disclose.

\vspace{0.2cm}

\noindent
\textbf{Data and code availability}. The synthetic data used in the simulation study were generated using the procedures described in Section \ref{sec4}. The real dataset analyzed in Section \ref{sec5} is publicly available from the authors of \cite{lundy89}. The \texttt{R} code implementing the proposed FE3KM algorithm is available in the Supplementary Material.

\bibliographystyle{apalike}
\setlength{\bibsep}{0.1pt}
\bibliography{bibliography}

\clearpage

\appendix

\section{Example of generated dataset}\label{app:example}

\begin{figure}[ht!]
\centering
\includegraphics[scale=0.35]{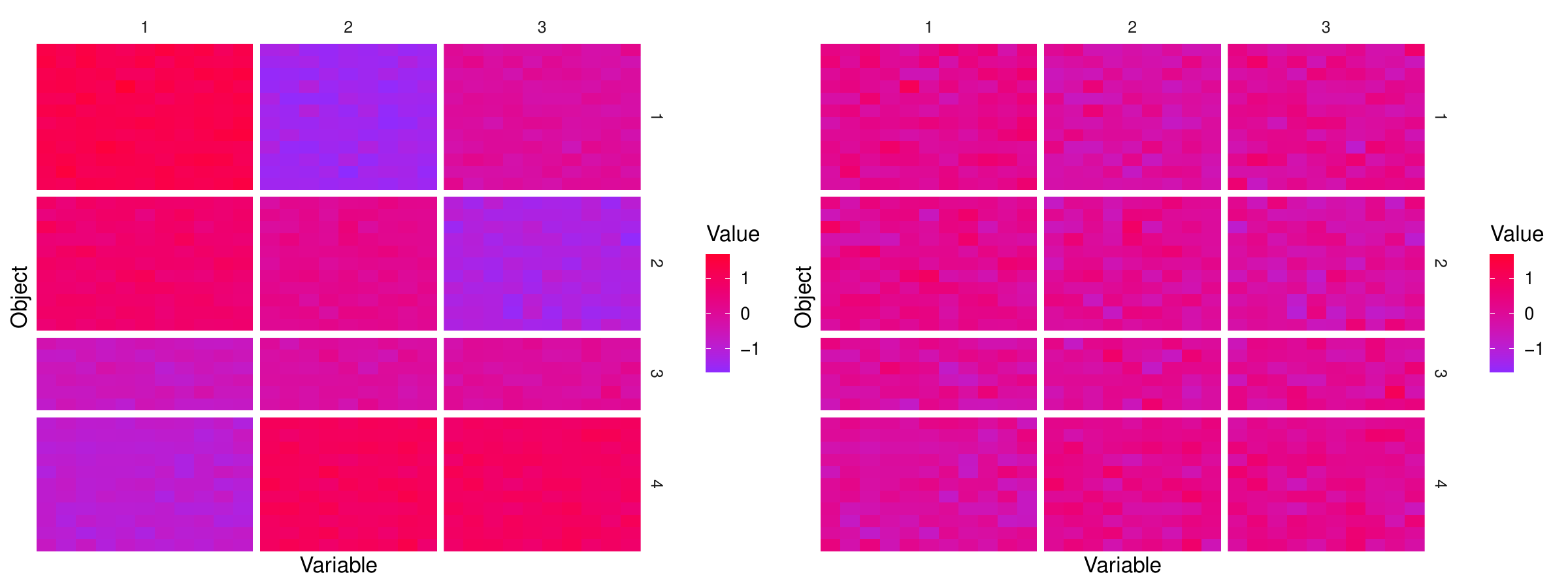}

\caption{Example of simulated sample for the Small-Small scenario, for low (left) and high (right) error. Object-Variable clusters heatmaps are obtained averaging over the occasions in Cluster 2}
\label{fig1.0}       
\end{figure}

\section{Lambda mis-specification: variables and occasions}\label{app:mis}

\begin{figure}[ht!]
\centering
\includegraphics[scale=0.35]{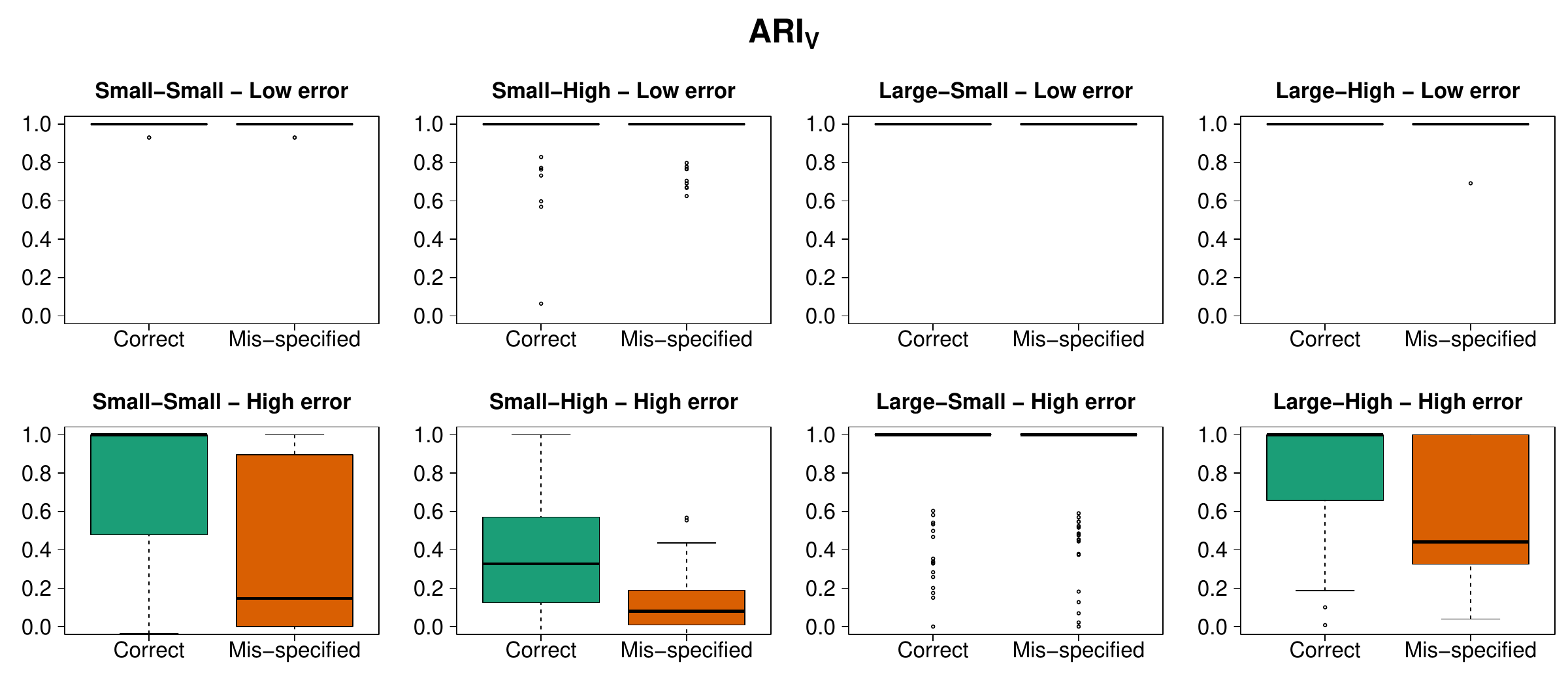}
\caption{Boxplots of ARI$_\mathbf{V}$ values under calibrated and mis-specified $\lambda$}
\label{boxplotV}
\end{figure}

\begin{figure}[ht!]
\centering
\includegraphics[scale=0.35]{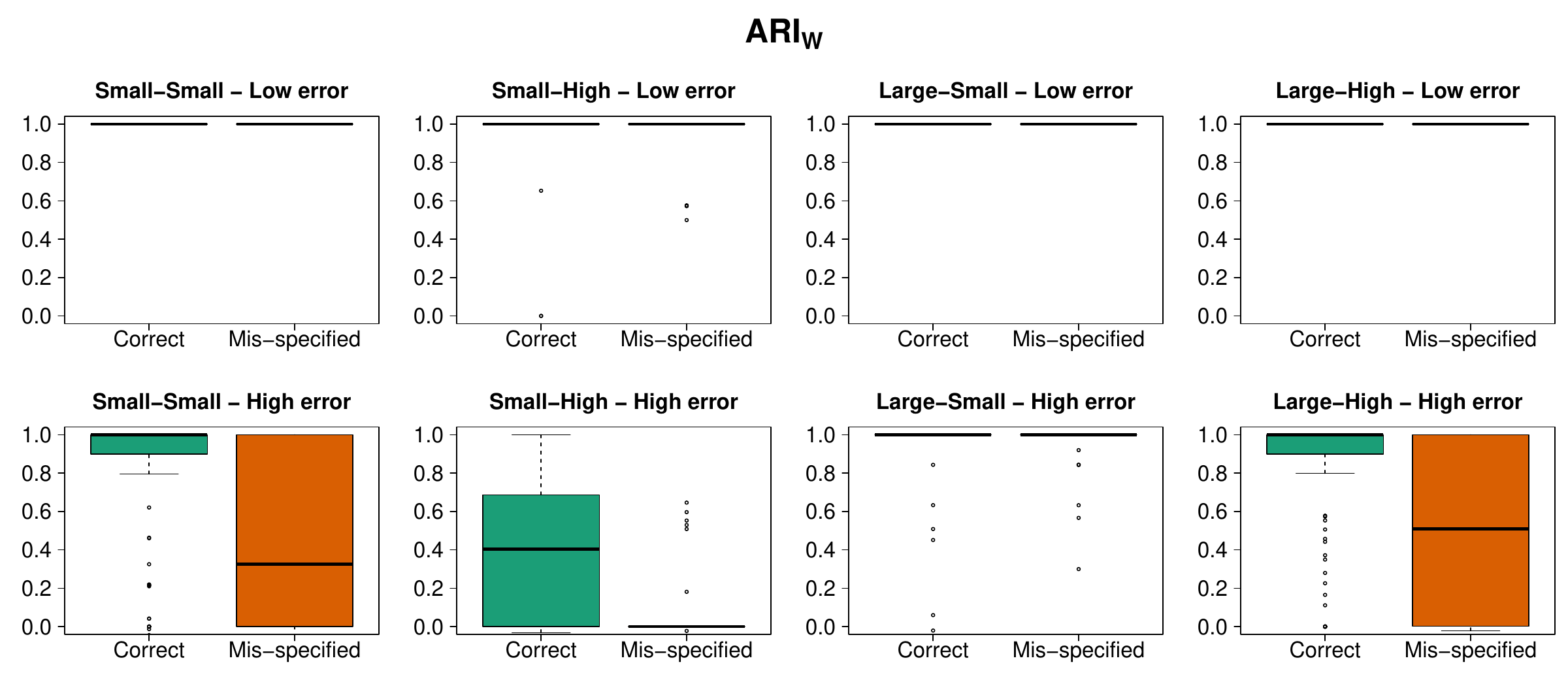}
\caption{Boxplots of ARI$_\mathbf{W}$ values under calibrated and mis-specified $\lambda$}
\label{boxplotW}
\end{figure}

\newpage

\section{Comparison with tandem and independent procedures: variables and occasions}\label{app:comparison}
 
\begin{figure}[ht!]
\centering
\includegraphics[width=\textwidth]{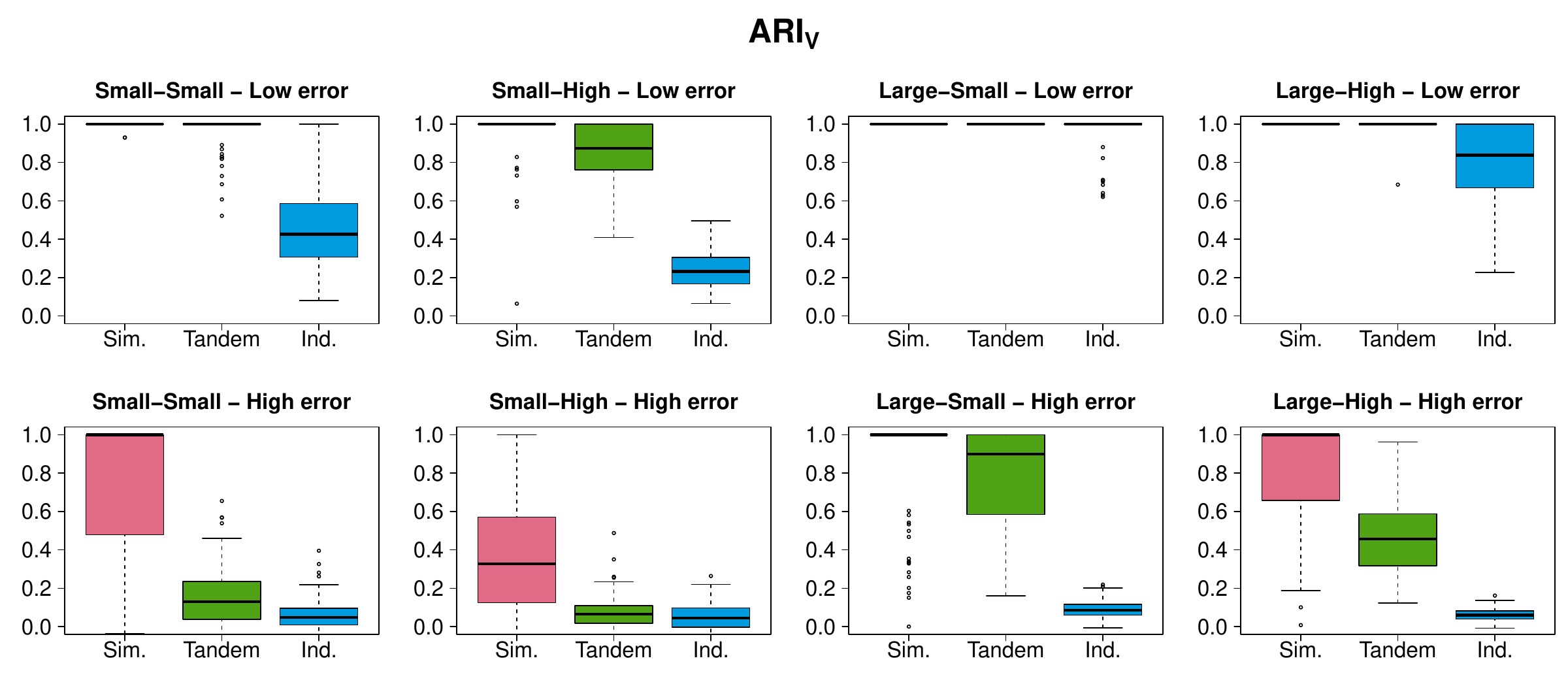}
\caption{Boxplots of ARI$_{\mathbf{V}}$ for the simultaneous (Sim.), tandem (Tandem) and independent (Ind.) procedures. Panels are labeled as data size--number of clusters -- noise level}
\label{fig:cmp_V}
\end{figure}
 
\begin{figure}[ht!]
\centering
\includegraphics[width=\textwidth]{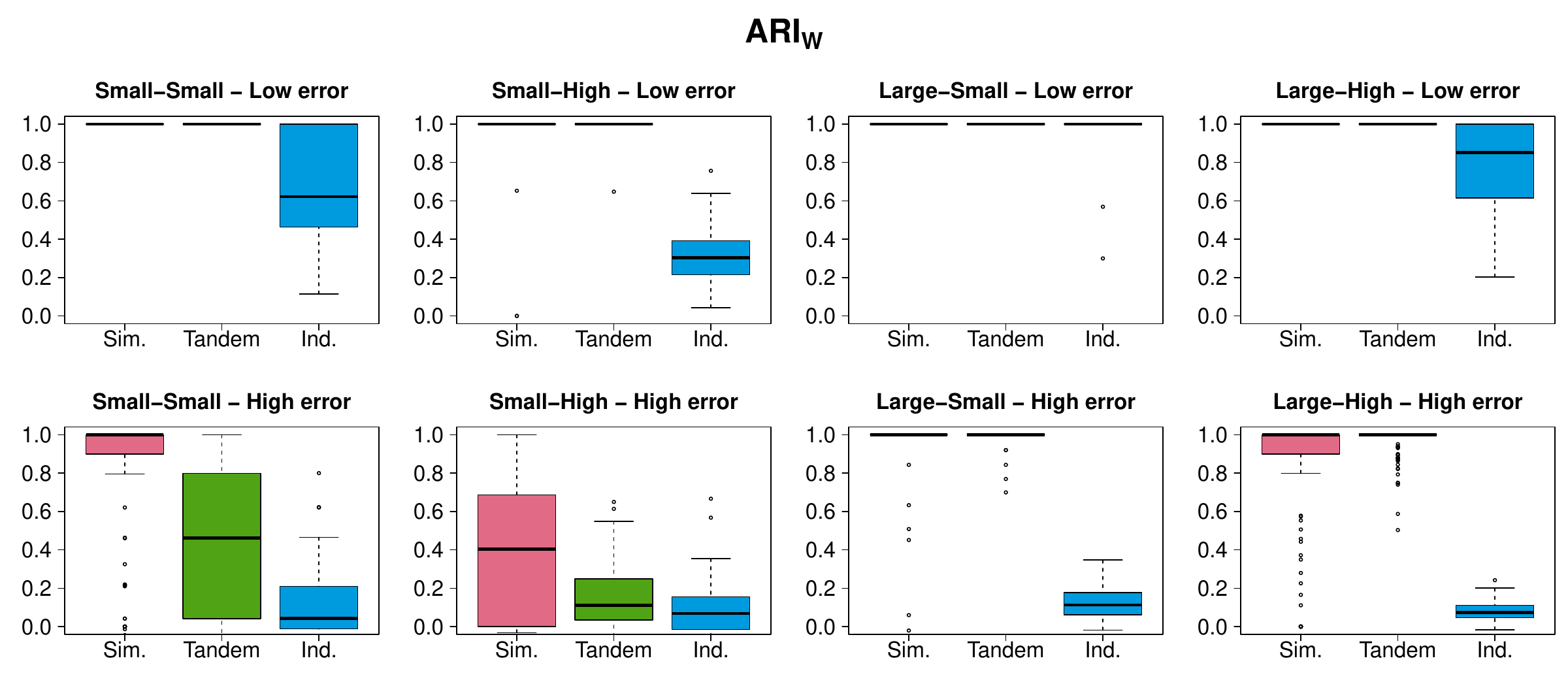}
\caption{Boxplots of ARI$_{\mathbf{W}}$ for the simultaneous (Sim.), tandem (Tandem) and independent (Ind.) procedures. Panels are labeled as data size--number of clusters -- noise level}
\label{fig:cmp_W}
\end{figure}

\newpage

\section{Xie--Beni values for the grid search of TV data application}\label{app:xb}

\newcommand{\xe}[2]{$#1{\times}10^{#2}$}

\begin{table}[ht!]
\caption{XB values for $P,Q,R \in \{2,\ldots,6\}$. Each panel corresponds to a fixed number of student clusters $R$; rows correspond to $P$ (scale clusters), columns to $Q$ (TV show clusters). The global minimum, attained at $(P,Q,R)=(3,6,2)$, is shown in bold}
\label{tab:XB}
\centering
\scriptsize
\setlength{\tabcolsep}{3pt}

\begin{minipage}[t]{0.48\textwidth}
\centering
\textbf{$R=2$}\\[2pt]
\begin{tabular}{c|ccccc}
\hline
$P \backslash Q$ & 2 & 3 & 4 & 5 & 6 \\
\hline
2 & 10.2 & 10.1 & 9.2 & 8.4 & 48.2 \\
3 & 14.9 & 15.8 & 9.4 & 8.1 & \textbf{7.7} \\
4 & \xe{1.4}{6} & \xe{6.4}{3} & \xe{3.5}{6} & \xe{1.7}{3} & \xe{4.2}{3} \\
5 & \xe{4.8}{3} & \xe{1.6}{5} & \xe{1.6}{3} & \xe{2.2}{4} & \xe{7.1}{3} \\
6 & \xe{4.2}{7} & \xe{1.4}{5} & \xe{8.6}{5} & \xe{1.4}{4} & \xe{6.1}{14} \\
\hline
\end{tabular}
\end{minipage}%
\hfill
\begin{minipage}[t]{0.48\textwidth}
\centering
\textbf{$R=3$}\\[2pt]
\begin{tabular}{c|ccccc}
\hline
$P \backslash Q$ & 2 & 3 & 4 & 5 & 6 \\
\hline
2 & \xe{5.1}{5} & \xe{1.2}{5} & \xe{1.1}{13} & \xe{3.2}{6} & \xe{4.6}{6} \\
3 & \xe{8.7}{5} & \xe{7.3}{5} & \xe{1.3}{5} & \xe{9.4}{9} & \xe{1.8}{7} \\
4 & \xe{3.2}{6} & \xe{4.0}{5} & \xe{4.1}{7} & \xe{1.8}{6} & \xe{1.3}{4} \\
5 & \xe{7.4}{7} & \xe{1.7}{5} & \xe{3.2}{9} & \xe{5.8}{8} & \xe{1.5}{4} \\
6 & \xe{1.9}{10} & \xe{1.9}{7} & \xe{6.0}{10} & \xe{5.7}{4} & \xe{2.2}{6} \\
\hline
\end{tabular}
\end{minipage}

\vspace{0.4cm}

\begin{minipage}[t]{0.48\textwidth}
\centering
\textbf{$R=4$}\\[2pt]
\begin{tabular}{c|ccccc}
\hline
$P \backslash Q$ & 2 & 3 & 4 & 5 & 6 \\
\hline
2 & \xe{1.6}{9} & \xe{1.4}{6} & \xe{3.1}{10} & \xe{6.2}{8} & \xe{1.0}{6} \\
3 & \xe{5.4}{8} & \xe{5.9}{7} & \xe{8.1}{6} & \xe{1.7}{7} & \xe{4.3}{6} \\
4 & \xe{5.3}{7} & \xe{2.3}{6} & \xe{8.2}{6} & \xe{9.1}{6} & \xe{1.8}{6} \\
5 & \xe{5.6}{6} & \xe{4.1}{6} & \xe{1.6}{8} & \xe{4.0}{7} & \xe{7.7}{4} \\
6 & \xe{3.5}{9} & \xe{8.8}{7} & \xe{1.5}{6} & \xe{1.4}{6} & \xe{5.5}{9} \\
\hline
\end{tabular}
\end{minipage}%
\hfill
\begin{minipage}[t]{0.48\textwidth}
\centering
\textbf{$R=5$}\\[2pt]
\begin{tabular}{c|ccccc}
\hline
$P \backslash Q$ & 2 & 3 & 4 & 5 & 6 \\
\hline
2 & \xe{8.5}{4} & \xe{4.5}{11} & \xe{1.1}{7} & \xe{2.7}{5} & \xe{1.2}{6} \\
3 & \xe{2.6}{6} & \xe{5.7}{6} & \xe{1.3}{11} & \xe{4.0}{5} & \xe{4.8}{5} \\
4 & \xe{5.5}{6} & \xe{5.9}{5} & \xe{1.5}{5} & \xe{1.3}{6} & \xe{9.0}{5} \\
5 & \xe{2.8}{9} & \xe{1.2}{7} & \xe{2.1}{9} & \xe{4.8}{11} & \xe{8.1}{7} \\
6 & \xe{4.6}{5} & \xe{1.2}{7} & \xe{1.5}{9} & \xe{7.2}{6} & \xe{6.4}{5} \\
\hline
\end{tabular}
\end{minipage}

\vspace{0.4cm}

\begin{minipage}[t]{0.48\textwidth}
\centering
\textbf{$R=6$}\\[2pt]
\begin{tabular}{c|ccccc}
\hline
$P \backslash Q$ & 2 & 3 & 4 & 5 & 6 \\
\hline
2 & \xe{1.7}{5} & \xe{2.7}{5} & \xe{1.3}{7} & \xe{1.2}{8} & \xe{1.5}{12} \\
3 & \xe{1.5}{7} & \xe{2.3}{6} & \xe{2.5}{6} & \xe{9.1}{8} & \xe{1.7}{6} \\
4 & \xe{4.0}{5} & \xe{8.1}{5} & \xe{5.0}{9} & \xe{2.1}{5} & \xe{1.5}{4} \\
5 & \xe{2.1}{7} & \xe{1.5}{11} & \xe{1.4}{6} & \xe{3.2}{5} & \xe{1.1}{6} \\
6 & \xe{4.4}{9} & \xe{2.0}{7} & \xe{4.3}{7} & \xe{6.8}{4} & \xe{1.7}{5} \\
\hline
\end{tabular}
\end{minipage}
\end{table}

\end{document}